\documentclass[reprint,onecolumn,showpacs,nofootinbib,prd,amssymb]{revtex4}
\usepackage{latexsym}
\usepackage{graphicx,epsf, epsfig, amssymb, amsmath}

\usepackage[usenames,dvipsnames]{xcolor}

\usepackage{relsize}
\usepackage{dsfont}
\usepackage{todonotes}
\usepackage{textcomp}
\usepackage{float}
\usepackage{color}
\usepackage{mathrsfs}

\newcommand{\aap}{Astron.\ Astrophys.}
\newcommand{\mnras}{Mon.\ Not.\ R.\ Astron.\ Soc.}
\newcommand{\physrep}{Phys.\ Rep.}
\newcommand{\apjl}{Astrophys.\ J.\ Lett.}
\newcommand{\apjs}{Astrophys.\ J.\ Suppl.\ Ser.}

\newcommand{\apss}{Astrophys.\ Space Sci.}
\newcommand{\ssr}{Sp. Sci. Rev.}

\newcommand{\const}{\mathrm{const}}
\newcommand{\Tg}{\widetilde{T}}
\newcommand{\diff}{\mathrm{d}}
\newcommand{\pd}[2]{\frac{\partial #1}{\partial #2}}
\DeclareMathOperator{\rot}{curl}
\DeclareMathOperator{\diver}{div}
\renewcommand{\vec}[1]{\pmb{#1}}
\newcommand{\vnabla}{\vec{\nabla}}

\newcommand{\PLeg}{\hat{\bf P}}
\newcommand{\pind}{{(p)}}
\newcommand{\tind}{{(t)}}
\newcommand{\grsh}{\Delta{\!\!}^*}

\begin{document}

\title{Fast magnetic field evolution in neutron stars: 
the key role of magnetically induced fluid motions in the core}

\author{D.~D.~Ofengeim \& M.~E.~Gusakov} 
\affiliation{Ioffe Institute, Polytekhnicheskaya 26, 194021 Saint Petersburg, Russia}

\begin{abstract} 
In [Gusakov et al.\ PRD, 96, 103012, (2017)],
we proposed a self-consistent method to study the quasistationary evolution 
of the magnetic field in neutron-star cores. 
Here we apply it to calculate the instantaneous particle velocities 
and other parameters of interest, 
which are fixed by specifying the magnetic field configuration. 
Interestingly, we found that the magnetic field can lead
to generation of a macroscopic fluid motion with the velocity, significantly exceeding the diffusion particle velocities.
This result calls into question the standard view 
on the magnetic field evolution in neutron stars 
and suggests a new, shorter timescale for such evolution.
\end{abstract}

\date{\today}



\maketitle

\section{Introduction}
\label{sec:intro}

One of the most interesting problems in the neutron star (NS) physics 
concerns
the evolutionary relationships  
between various classes of NSs with often
very different surface magnetic fields ranging from,
e.g., $10^8-10^{10}$~G in millisecond pulsars to $\sim10^{12}$~G 
in ordinary radio pulsars and up to $\sim 10^{15}$~G
in magnetars.
What makes these fields so different?
The complete answer is still unknown 
in spite of a lot of effort devoted to this problem in the past.
Obviously, the magnetic field at the stellar surface  
should be related somehow to 
that
in NS interiors. 
So far, most of the research has been focused on the magnetic 
field evolution in the crust
(e.g., Refs.~\cite{jones88, su97, rg02, hr04, gcr_etal13, vigano_etal13, gc14, gwh16}).
The evolution in the core has been studied less intensively 
(see, e.g., Refs.~\cite{eprgv16,blb2017,papm17,crv2017}).
In part, this is because
such an analysis is substantially more complex, 
while the physics involved is not well understood,
and in part because 
of the widespread opinion 
(but see Refs.~\cite{jones06,blb2017}) 
that the
evolution in the crust proceeds on a shorter timescale than in the core.

This work continues our study of the magnetic field evolution in NS cores.
Recently, we have 
solved the following problem \cite{GKO17}.
Assume that there is an NS with some (specified) magnetic field $\vec{B}(t_0)$ 
at the initial moment of time $t=t_0$.
What will be the magnetic field in the next moment $t=t_0+\delta t$?
The answer to this question is provided by the Faraday's law, 
$\vec{B}(t_0+\delta t)=\vec{B}(t_0)- c \, \delta t \rot {\vec E}(t_0)$, 
so the problem reduces to finding the self-consistent electric field $\vec{E}$ at $t=t_0$.
The latter field depends, in turn, on perturbations of particle chemical potentials 
and various particle velocities induced by the magnetic field.
In Ref.~\cite{GKO17} we proposed a method to self-consistently calculate all these quantities. 
We also found that 
the previous works, 
exploring the same problem
(see, e.g., Refs.\ \cite{gr92,td96,hrv08, reisenegger09, hrv10,gjs11,bl16,papm17,crv2017}), 
have made some unjustified simplifications,
that can qualitatively change the results.

Here our aim is to
further develop the method of Ref.\ \cite{GKO17} 
and to demonstrate its ability 
to calculate 
all the 
ingredients necessary 
to follow
the evolution of $\vec{B}$ in time.
For this purpose, we adopt a model of an NS core
consisting of nonsuperfluid neutrons ($n$), nonsuperconducting protons ($p$), and electrons ($e$). 
Although very simplistic, this model may be adequate in describing hot magnetars 
with very high magnetic field, resulting in a partly (or fully) suppressed
nucleon superfluidity/superconductivity in their cores \cite{ss15,sxsc17}.
The main advantage of the model is that 
it allows for relatively simple and transparent calculations,
the disadvantage is that it ignores the effects of baryon superfluidity/superconductivity,
and thus is not directly applicable to cold NSs with relatively low magnetic field 
(e.g., radio- and millisecond pulsars); evolution of the magnetic field in such stars
will be considered by us elsewhere. 

The main result in the paper
consists in 
demonstrating 
that the relative particle velocities (generated by the magnetic field) 
can generally be much smaller than the velocity of the fluid as a whole.
This result is important since it
introduces a new small timescale into the problem of the magnetic field evolution in NSs. It also contradicts the general belief
that NS matter is motionless to a good approximation 
in the presence of the magnetic field (e.g., \cite{gr92,td96,bl16,papm17,crv2017}). 

The paper is organized as follows.
Sec.~\ref{sec:equ} discusses magnetohydrodynamic equations and various approximations
that were made to solve them.
In Sec.~\ref{sec:sol} we propose a detailed analytic solution to these equations.
Sec.~\ref{sec:figures} contains description of the adopted magnetic field models 
and our numerical results.
Finally, discussion and conclusions are presented in Sec.~\ref{disc}.

\section{General equations and quasistationary approximation}
\label{sec:equ}

In this section we formulate and briefly discuss the magnetohydrodynamic (MHD) equations
for nonsuperfluid and nonsuperconducting matter of NSs
composed of various particle species. 
The equation of state (EOS) is assumed to be relativistic, 
but the effects of general relativity will be disregarded for simplicity.
We also neglect (weak) thermal forces in the equations of motion for each particle species, 
as well as the effects of temperature on the EOS.  
With these simplifications,  
the system of MHD equations can be written as (e.g., \cite{ys91a,gr92,GKO17})
\begin{subequations}
\label{eq:MHDgen}
\begin{align}
&\pd{n_a}{t} + \diver \left( n_a \vec{u}_a \right) = \Delta\Gamma_a, \label{eq:MHDgen-cont} \\
&n_a \left[ \pd{}{t} + \left( \vec{u}_a\vec{\nabla} \right) \right]\left( \frac{\mu_a}{c^2}\vec{u}_a \right) = -n_a\vec{\nabla}\mu_a + e_a n_a \left( \vec{E} + \frac{\vec{u}_a}{c}\times\vec{B} \right) - \frac{\mu_a n_a}{c^2}\vec{\nabla}\phi - \sum_{b\ne a}J_{ab}\left( \vec{u}_a - \vec{u}_b \right), \label{eq:MHDgen-euler} \\
&\Delta\phi = \frac{4\pi G}{c^2}(P+\varepsilon), \label{eq:MHDgen-grav} \\
&\sum_a e_a n_a = 0, \label{eq:MHDgen-neutral} \\
&\frac{c}{4\pi}\rot \vec{B} = \sum_a e_a n_a \vec{u}_a = \vec{j}, \label{eq:MHDgen-amper} \\
&\pd{\vec{B}}{t} = - c\rot \vec{E}. \label{eq:MHDgen-farad}
\end{align}
\end{subequations}
Eqs.~(\ref{eq:MHDgen-cont}) and (\ref{eq:MHDgen-euler}) 
should be satisfied for each particle species ``$a$'' separately. 
Here $e_a$, $n_a$, $\mu_a$, and $\vec{u}_a$ are, respectively, 
the electric charge, number density, chemical potential, and velocity for particles ``$a$''; 
$\Delta\Gamma_a$ is the 
reaction
rate 
for 
species ``$a$'' 
due to nonequilibrium processes of particle mutual transformations; 
$J_{ab} = J_{ba}$ 
is the coefficient in the expression for the friction force, 
$-J_{ab}(\vec{u}_a-\vec{u}_b)$, describing friction between species $a$ and $b$;
it is related to the effective relaxation time, $\tau_{ab}$, by the formula \cite{ys91a}:
$J_{ab}= n_a \mu_a/(\tau_{ab} c^2)$. 
Further, $\phi$ is the Newtonian gravitational potential, 
$G$ and $c$ are, respectively, the gravitational constant and the speed of light; 
$P$ and $\varepsilon = \rho c^2$ are the total pressure and energy density 
(proportional to the total density of matter $\rho$); 
$\vec{j}$, $\vec{E}$, and $\vec{B}$ are the electric current density, electric, and magnetic fields, respectively. 

The thermodynamic parameters appearing in Eqs.\ (\ref{eq:MHDgen}) are not all independent
and are related by the following formulas, valid in strongly degenerate matter,
\begin{subequations}
\label{eq:PEpsMuN}
\begin{align}
P + \varepsilon &= \sum_a n_a\mu_a, \label{eq:PEpsMuN-alg} \\
\diff P &= \sum_a n_a \diff \mu_a. \label{eq:PEpsMuN-diff}
\end{align}
\end{subequations}

In what follows we 
consider only small perturbations of the system (\ref{eq:MHDgen}) 
assuming that unperturbed state describes a nonmagnetized NS with
the core composed of neutrons ($n$), protons ($p$), and electrons ($e$) 
in full hydrostatic and thermodynamic equilibrium. 
Correspondingly, in the unperturbed star
$\vec{u}_a = 0$ and $\Delta\Gamma_a = 0$ for $a = n$, $p$, $e$,
while other parameters (e.g., the number densities $n_a$) are the only functions
of the radial coordinate $r$ (here and hereafter we make use of the spherical coordinates 
$r$, $\theta$, and $\varphi$).

Now let us slightly perturb the star by creating some small currents
that will generate the magnetic field in the system.
Following the standard approach of Ref.\ \cite{gr92} 
(see also, e.g., Refs.\ \cite{papm17,crv2017,GKO17,gjs11,bl16})	
we assume that the perturbed star is in the quasistationary equilibrium 
and is stable with respect to spontaneous reconfiguration of the magnetic field.
This allows us to
neglect time derivatives
in the Euler and the continuity equations, 
as discussed in detail below.
The generated magnetic field causes small deviations 
$\delta n_a$, $\delta\mu_a$, $\delta\vec{E}$, and $\delta\phi$ 
of the quantities $n_{a}$, $\mu_{a}$, $\vec{E}$, and $\phi$
from their unperturbed values $n_{a0}$, $\mu_{a0}$, $\vec{E}_0$, and $\phi_0$.
This launches two dissipation mechanisms: 
nonequilibrium reactions 
[see the reaction rates $\Delta\Gamma_a$ in Eq.~(\ref{eq:MHDgen-cont})] 
and diffusion with small but non-zero particle velocities $\vec{u}_a$
[see the corresponding friction forces, $-J_{ab}(\vec{u}_a-\vec{u}_b)$, 
in Eq.~(\ref{eq:MHDgen-euler})].
For $npe$-cores of NSs the only nonequilibrium reactions that generate
$\Delta\Gamma_a$ are the direct and modified Urca processes. 
The corresponding rates $\Delta\Gamma_n = - \Delta\Gamma_e = - \Delta\Gamma_p
\equiv \Delta\Gamma$ 
were calculated in Ref.\ \cite{Reis95}, where it was shown that
$\Delta\Gamma$ can be presented as the function of the chemical potential imbalance, 
$\Delta \mu \equiv \mu_e+\mu_p-\mu_n=\delta \mu_e+\delta\mu_p-\delta\mu_n$.%
%
\footnote{The latter equality follows from the fact that in beta-equilibrium
$\mu_{e0}+\mu_{p0}-\mu_{n0}=0$ (see, e.g., Ref.\ \cite{hpy07}).}
%
The 
expressions for these rates, as well as for the coefficients $J_{ab}$
are presented in Ref.\ \cite{GKO17}; 
we use the same expressions in all our calculations and
refer the interested reader to that reference for more details.
However, for subsequent presentation it is important to note
that the electron-neutron friction coefficient $J_{en}$ is much smaller 
than other coefficients, $J_{np},\, J_{ep} \gg J_{en}$, and therefore can be ignored.
In what follows $J_{np}$, $J_{ep}$, and $\Delta\Gamma$ 
are treated as given explicit functions of unperturbed number densities $n_a$ and temperature $T$.

Now our aim is to write down Eqs.~(\ref{eq:MHDgen}) for a perturbed star.
Let us, for example, consider equation of motion (\ref{eq:MHDgen-euler}) for a particle species $a$
\begin{multline}
\label{eq:eulerPert}
n_a \left[ \pd{}{t} + \left( \vec{u}_a \vnabla \right) \right]\left( \frac{\mu_a}{c^2}\vec{u}_a \right) = \delta n_a \left( -\vnabla\mu_a + e_a\vec{E} - \frac{\mu_a}{c^2}\vnabla\phi \right) - \frac{n_a}{c^2}\left( \delta\mu_a\vnabla\phi + \mu_a\vnabla\delta\phi \right) - \\
- n_a\vnabla\delta\mu_a + e_a n_a\left( \delta\vec{E} + \frac{\vec{u}_a}{c}\times\vec{B} \right) - \sum_{b\ne a}J_{ab}(\vec{u}_a - \vec{u}_b).
\end{multline}
Here and below to simplify notation,
$n_a$, $\mu_a$, $\vec{E}$, and $\phi$ stand for the unperturbed quantities. 
Note that even in the full equilibrium $\vec{E}\ne 0$, since a self-consistent electric field
is generated in order to preserve the quasineutrality condition, $n_e=n_p$ 
(see Eq.\ \ref{eq:MHDgen-neutral}).
The first term on the right-hand side of this equation vanishes because 
$-\vnabla\mu_a + e_a\vec{E} - \mu_a\vnabla\phi/c^2=0$ for a star in hydrostatic equilibrium.
Eq.~(\ref{eq:eulerPert}) 
(and other equations in the system~\ref{eq:MHDgen})
can be further simplified provided that
we make some reasonable approximations, 
which were discussed in detail in Refs.\ \cite{gr92,papm17,GKO17}.

\noindent (i) We neglect perturbations of gravitational potential $\delta\phi$ 
(Cowling approximation \cite{cow41}) and, moreover, 
the term $n_a \delta\mu_a\vnabla\phi/c^2$, 
since it have to be much smaller than $n_a\vec{\nabla}\delta\mu_a$ for a Newtonian star 
for which Eq.\ (\ref{eq:eulerPert}) is written.%
%
\footnote{In principle, it is easy to take this term into account,
which does not affect our results much.}
%
Thus only the last three terms survive in the right-hand side of Eq.~(\ref{eq:eulerPert}).

\noindent (ii) We use a quasistationary approximation. 
In our problem the magnetic field $\vec{B}$ 
is the 
only perturbing factor that drives the system out of the equilibrium.
Let $\tau_B$ and $L_B$ be the typical timescale and lengthscale of $\vec{B}$ evolution. 
Then one can estimate $\partial/\partial t \sim 1/\tau_B$, 
$\nabla\sim\diver\sim\rot \sim 1/L_B$, 
and $u_a \sim L_B/\tau_B$ in Eqs.~(\ref{eq:MHDgen}) and (\ref{eq:eulerPert}). 
Correspondingly, the left-hand side of Eq.~(\ref{eq:eulerPert}) can be estimated as
\begin{equation}
\label{eq:estimEuler-Left}
n_a \left[ \pd{}{t} + (u_a\nabla)\right] \left( \frac{\mu_a}{c^2}u_a \right) \sim n_a \frac{\mu_a u_a}{c^2\tau_B} \sim n_a \frac{\mu_a}{L_B}\frac{L_B}{c\tau_B}\frac{u_a}{c}\sim n_a \frac{\mu_a}{L_B} \frac{u_a^2}{c^2}.
\end{equation}
In turn, the terms in the right-hand side of Eq.~(\ref{eq:eulerPert}) are of the order of
$n_a \mu_a B^2/(8\pi P L_B)$ \cite{GKO17}, so that the left-hand side
terms can be safely neglected provided that 
$(u_a^2/c^2)(8\pi P)/B^2 \ll 1$ 
which is always the case 
for the typical NS conditions. 
Similar estimate can be made for the continuity equation (\ref{eq:MHDgen-cont})
where the time derivatives can also be omitted provided that $B^2/(8 \pi P)\ll 1$ \cite{gr92,GKO17}.

\noindent (iii) The system is assumed to be axisymmetric, 
i.e., 
$\vec{B}$ and all the perturbations depend exclusively on $r$ and $\theta$.

Using the above approximations 
the system of equations, 
describing $\vec{B}$-evolution,
takes the form (see also Ref.~\cite{GKO17})
\begin{subequations}
\label{eq:MHDquas}
\begin{align}
&\diver (n_n \vec{u}_n) = -\diver (n_e \vec{u}_p) = - \diver(n_e \vec{u}_e) = \Delta\Gamma, \label{eq:MHDquas-cont}\\
&n_n\vec{\nabla}\delta\mu_n + J_{np}(\vec{u}_n - \vec{u}_p) = 0, \label{eq:MHDquas-nEuler}\\
&n_e\vec{\nabla}\delta\mu_e + J_{ep}(\vec{u}_e - \vec{u}_p) = -e n_e \left( \delta\vec{E} + \frac{\vec{u}_e}{c}\times\vec{B} \right), \label{eq:MHDquas-eEuler}\\
&n_n\vec{\nabla}\delta\mu_n + n_e \vec{\nabla}(\delta\mu_p + \delta\mu_e) 
= \vec{f}_A, 
\label{eq:MHDquas-totEuler}\\
&\vec{j} = e n_e (\vec{u}_p - \vec{u}_e), \label{eq:MHDquas-amper} \\ 
&\pd{\vec{B}}{t} = - c\rot \delta\vec{E}. \label{eq:MHDquas-farad}
\end{align}
\end{subequations}
where 
$e=e_p$ is the elementary charge and $\vec{f}_{A}$ is the Ampere force,
\begin{equation}
\vec{f}_A = \frac{1}{c} \, [\vec{j}\times\vec{B}].
\label{fA}
\end{equation}
In the Euler Eqs.~(\ref{eq:MHDquas-nEuler})--(\ref{eq:MHDquas-totEuler}) 
the equation for protons is replaced with 
the total force balance Eq.~(\ref{eq:MHDquas-totEuler}),
which is a sum of Euler equations over all particle species.
Note that Eq.~(\ref{eq:MHDquas-totEuler}) ensures that the system is in hydrostatic
equilibrium during its (quasistationary) evolution.
The quasineutrality condition (\ref{eq:MHDgen-neutral}) 
is accounted for by substituting $n_p$ for $n_e$ in Eqs.\ (\ref{eq:MHDquas}); 
the unperturbed electric field does not appear
in Eq.~(\ref{eq:MHDquas-farad}) since it is purely potential. 

Now our aim is to find the particle velocities $\vec{u}_a$ from Eqs.~(\ref{eq:MHDquas}), 
induced by the presence of an axisymmetric magnetic field (which is assumed to be specified). 
As a by-product of this calculation, we will also find $\delta \vec{E}$ 
and, consequently, $\partial\vec{B}/\partial t$ (see Eq.\ \ref{eq:MHDquas-farad}).
This will open an exciting possibility to follow the quasistationary evolution of the magnetic field 
by simply iterating Eq.\ (\ref{eq:MHDquas-farad}) in time.

\section{Solution to Eqs.\ (\ref{eq:MHDquas})}
\label{sec:sol}

First, let us define the baryon velocity $\vec{U}_b$ 
and the diffusion velocities $\vec{w}_a$ according to the equations
\begin{equation}
\label{eq:UWdef}
\vec{U}_b \equiv \frac{n_n}{n_b}\vec{u}_n + \frac{n_p}{n_b}\vec{u}_p
=\frac{n_n}{n_b}\vec{u}_n + \frac{n_e}{n_b}\vec{u}_p , 
\qquad \vec{w}_a \equiv \vec{u}_a - \vec{U}_b,
\end{equation}
where $n_b \equiv n_n + n_p=n_n+n_e$ is the baryon number density. 
Second, let us introduce the electric field ${\pmb E}_{\rm com}$ in the frame, 
locally comoving with the baryons, that is
\begin{equation}
\label{eq:EcomDef}
\delta\vec{E}_\text{com} = \delta\vec{E} + \frac{\vec{U}_b}{c}\times\vec{B}.
\end{equation}
We assume that the (axisymmetric) instantaneous magnetic field $\vec{B}$ 
is specified, 
i.e., we know the function ${\pmb B}(r,\, \theta)$ at some moment of time
(but the time dependence of ${\pmb B}$ is unknown!). 
In view of Eqs.~(\ref{eq:MHDgen-amper}) and (\ref{fA})
this means that
the vectors $\vec{j}$ and $\vec{f}_A$ are also specified. 
In what follows it will be convenient to rewrite the Faraday's law in terms 
of the quantities $\delta\vec{E}_\text{com}$ and $\vec{U}_b$. 
Plugging Eq.~(\ref{eq:EcomDef}) into (\ref{eq:MHDquas-farad}), we get
\begin{equation}
\label{eq:faradCom}
\pd{\vec{B}}{t} = - c \rot\delta\vec{E}_\text{com} + \rot\left( \vec{U}_b\times\vec{B} \right).
\end{equation}
Thus, in order to trace the evolution of the magnetic field,  
one has to determine the comoving electric field and the baryon velocity. 
Note that in Ref.\ \cite{GKO17} we used the net flow velocity 
$\vec{U} = \sum_a \mu_a n_a \vec{u}_a/(P+\varepsilon)$ instead of $\vec{U}_{b}$. 
This approach is equivalent to that employed here since 
$\vec{U}_b = \vec{U} + \vec{j}\mu_e/[e (P+\varepsilon)]$.

\subsection{Finding $\vec{w}_a$ and $\delta \vec{E}_{\rm com}$}
\label{sec:sol-com}

Let us assume for a while that the quantities $\delta\mu_a$ are known. 
Then we can easily find the vectors $\vec{w}_a$ 
(which are the velocities of particle species $a$ 
in the frame comoving with baryons, $\vec{U}_b=0$). 
Using Eqs.~(\ref{eq:MHDquas-nEuler}), (\ref{eq:MHDquas-amper}), and (\ref{eq:UWdef}), one has
\begin{subequations}
\label{eq:Wsys}
\begin{align}
J_{np}\left( \vec{w}_n - \vec{w}_p \right) &= - n_n\vec{\nabla}\delta\mu_n, \label{eq:Wsys-np}\\
e n_e \left( \vec{w}_p - \vec{w}_e \right) &= \vec{j}, \label{eq:Wsys-pe}\\
n_n \vec{w}_n + n_e \vec{w}_p &= 0. \label{eq:Wsys-tot}
\end{align}
\end{subequations}
The solution to this system is
\begin{subequations}
\label{eq:Wsol}
\begin{align}
\vec{w}_n &= - \frac{n_e n_n}{n_b J_{np}}\vec{\nabla}\delta\mu_n, \label{eq:Wsol-n}\\
\vec{w}_p &= \frac{n_n^2}{n_b J_{np}}\vec{\nabla}\delta\mu_n, \label{eq:Wsol-p}\\
\vec{w}_e &= \frac{n_n^2}{n_b J_{np}}\vec{\nabla}\delta\mu_n - \frac{\vec{j}}{e n_e}. \label{eq:Wsol-e}
\end{align}
\end{subequations}
Eqs.~(\ref{eq:MHDquas-eEuler}), (\ref{eq:EcomDef}), and (\ref{eq:Wsol-e}) give
\begin{equation}
\label{eq:EcomSol}
\delta\vec{E}_\text{com} = -\frac{\vnabla\delta\mu_e}{e} + \frac{J_{ep}}{e^2 n_e^2}\vec{j} - \frac{n_n^2}{c n_b J_{np}}\vnabla\delta\mu_n\times\vec{B} + \frac{\vec{f}_A}{e n_e}.
\end{equation}
Actually, 
in order to follow the magnetic field evolution,
we are interested in the quantity  
${\rm curl} \, \delta\vec{E}_\text{com}$ 
rather than in $\delta\vec{E}_\text{com}$ 
(see Eq.~\ref{eq:faradCom}). 
This means, in particular, that there is no need 
in calculation of
the term 
$\vec{\nabla}\delta\mu_e/e$ in Eq.\ (\ref{eq:EcomSol}).

Note that Eqs.~(\ref{eq:Wsol}) and (\ref{eq:EcomSol}) can also be obtained 
using a diffusion tensor from Ref.\ \cite{ys91a}, 
but this way of derivation is more sophisticated. 
It leads to the generalized form of the Ohm's law 
\cite{su95}, 
which is equivalent to Eqs.~(\ref{eq:Wsol}) and (\ref{eq:EcomSol}).

\subsection{Finding the poloidal component of the baryon velocity $\vec{U}_b$}
\label{sec:sol-Up}

First, let us note that, 
according to Eqs.~(\ref{eq:MHDquas-cont}) and (\ref{eq:UWdef}), 
the baryon velocity is purely solenoidal,
\begin{equation}
\label{eq:nbcont}
\diver\left( n_b \vec{U}_b \right) = 0.
\end{equation}
Recalling that the system is axisymmetric, 
the expression for the poloidal component of the baryon velocity can be presented as
\begin{equation}
\label{eq:ups-Intro}
n_b \vec{U}_b^\pind = \rot\left[ \Upsilon(r,\theta)\vnabla\varphi \right] = \vnabla\Upsilon\times\vnabla\varphi,
\end{equation}
where $\vnabla\varphi = \vec{e}_\varphi/(r\sin\theta)$; 
the superscript ``$\pind$'' denotes the poloidal component of a vector field;
and $\Upsilon$ is an arbitrary function of $r$ and $\theta$.
This decomposition is widely used for poloidal magnetic fields 
(see, e.g., Ref.\ \cite{Goed2010} and Sec.~\ref{sec:sol-Uphi}). 
Then, plugging Eqs.~(\ref{eq:UWdef}) and (\ref{eq:ups-Intro})
into Eq.~(\ref{eq:MHDquas-cont}) for protons, we have 
\begin{equation}
\label{eq:eq:pcont+ups}
\diver\left( n_e \vec{u}_p \right) = \diver\left( \frac{n_e}{n_b}\vnabla\Upsilon\times\vnabla\varphi \right) + \diver\left( n_e \vec{w}_p \right) = \vnabla\left( \frac{n_e}{n_b} \right)\cdot\left( \vnabla\Upsilon\times\vnabla\varphi \right) + \diver\left( n_e \vec{w}_p \right) = - \Delta\Gamma.
\end{equation}
Employing Eq.~(\ref{eq:Wsol-p}), 
we arrive at
\begin{equation}
\label{eq:upsEq-vec}
\left[ \vnabla\varphi\times\vnabla\left( \frac{n_e}{n_b} \right) \right] \cdot \vnabla\Upsilon = -\diver\left( \frac{n_e n_n^2}{n_b J_{np}}
\vnabla\delta\mu_n \right) - \Delta\Gamma.
\end{equation}
Since $\partial n_a/\partial\theta = \partial n_a/\partial\varphi = 0$, this equation takes the form
\begin{equation}
\label{eq:upsEq-scal}
\pd{\Upsilon}{\theta} = n_b r^2 V(r,\theta) \sin\theta,
\end{equation}
where  
\begin{equation}
\label{eq:Vdef}
V = \frac{n_b}{n_b' n_e - n_e' n_b}\left[ \diver\left( \frac{n_e n_n^2}{n_b J_{np}}\vec{\nabla}\delta\mu_n \right) + \Delta\Gamma \right],
\end{equation}
and a prime sign $(')$ stands for $\diff/\diff r$. 
The solution to Eq.~(\ref{eq:upsEq-scal}) is
\begin{equation}
\label{eq:upsSol}
\Upsilon(r,\theta) = n_b r^2 \int_0^\theta V(r,\tilde{\theta})\sin\tilde{\theta}\diff\tilde{\theta} + \xi(r),
\end{equation}
where $\xi(r)$ is an arbitrary function. 
Plugging this solution into Eq.~(\ref{eq:ups-Intro}), 
we find an expression for the poloidal component of the baryon velocity, 
which contains the term $\xi'(r)/(r\sin\theta)$. 
This velocity must be finite at the symmetry axis, $\theta=0$, 
which
leads to the condition
$\xi'(r)\equiv 0$. 
The final expression for
$\vec{U}_b^\pind = (U_{br}, U_{b\theta}, 0)$ 
takes the form
\begin{equation}
\label{eq:Upfin}
U_{br} = V, \qquad U_{b\theta} = - \frac{1}{n_b r \sin\theta}\pd{}{r}\left( n_b r^2 \int_0^\theta V \sin\tilde{\theta}\diff\tilde{\theta} \right).
\end{equation}
Note that, following the method described in our earlier paper \cite{GKO17}, 
one would obtain similar expressions for the net flow velocity $\vec{U}$. 
Using them to derive the baryon velocity, one would arrive at Eq.~(\ref{eq:Upfin}),
as it should be.
Now we have the explicit expressions (\ref{eq:Vdef}) and (\ref{eq:Upfin}) 
for the velocity
$\vec{U}_b^\pind$. 
But they still depend on the unknown perturbations of chemical potentials, $\delta\mu_a$. 
Our aim is to find them.

\subsection{Perturbations of chemical potentials}
\label{sec:sol-mu}

Let us consider Eq.~(\ref{eq:MHDquas-totEuler}). 
Introducing the chemical potential imbalance, 
$\Delta\mu = \delta\mu_p+\delta\mu_e-\delta\mu_n$,
it can be rewritten as
\begin{equation}
\label{eq:fAbalance}
n_b \vec{\nabla}\delta\mu_n + n_e \vec{\nabla}\Delta\mu = \vec{f}_A.
\end{equation}
Its analytical solution is proposed in Ref.~\cite{GKO17}. 
Here we just briefly outline it. 
First, we remind that our problem is axisymmetric, which means that the  
$\varphi$-component of the Ampere force must vanish, $f_{A \varphi}=0$.
Second, since $\partial n_a/\partial\theta = 0$, Eq.~(\ref{eq:fAbalance}) yields
\begin{subequations}
\label{eq:naToSolve}
\begin{align}
\pd{}{\theta}\left( n_b \delta\mu_n + n_e \Delta\mu \right) &= r f_{A\theta}, \label{eq:naToSolve-th}\\
n_b' \delta\mu_n + n_e' \Delta\mu &= \pd{}{r}\left( n_b \delta\mu_n + n_e \Delta\mu \right) - f_{Ar}. \label{eq:naToSolve-r}
\end{align}
\end{subequations}
Integrating Eq.~(\ref{eq:naToSolve-th}) with respect to $\theta$ and substituting 
the result into Eq.~(\ref{eq:naToSolve-r}), we get
\begin{subequations}
\label{eq:naAlgProb}
\begin{align}
n_b \delta\mu_n + n_e \Delta\mu & = r \int_0^\theta f_{A\theta}\diff\tilde{\theta} + \zeta(r), \label{eq:naAlgProb-up}\\
n_b' \delta\mu_n + n_e' \Delta\mu & = \pd{}{r}\left( r \int_0^\theta f_{A\theta}\diff\tilde{\theta} \right) - f_{Ar} + \zeta'(r), \label{eq:naAlgProb-down}
\end{align}
\end{subequations}
where $\zeta(r)$ is an arbitrary function. 
In what follows it will be convenient to use the operator $\PLeg_l$ 
that extracts the $l$'th Legendre component of its argument,
\begin{equation}
\label{eq:PlegDef}
\PLeg_l(\cdot) \equiv \frac{2l+1}{2}\int_0^\pi (\cdot) P_l(\cos\theta) \sin\theta\diff\theta,
\end{equation}
where $P_l({\rm cos} \theta)$ is the $l$'th Legendre polynomial.
Then for $l\geqslant 1$ we have
\begin{equation}
\label{eq:DmudmuHighL}
\begin{pmatrix}
\PLeg_l\delta\mu_n \\
\PLeg_l\Delta\mu
\end{pmatrix}
=
\begin{pmatrix}
n_e' & n_b' \\
n_e & n_b
\end{pmatrix}^{-1} 
\begin{pmatrix}
\left[ r \PLeg_l \int_0^\theta f_{A\theta}\diff\theta \right]' - \PLeg_l f_{Ar}\\
r \PLeg_l \int_0^\theta f_{A\theta}\diff\theta
\end{pmatrix}, 
\qquad  l \geqslant 1.
\end{equation}
Notice that for $l\geqslant 1$ $\PLeg_l\delta\mu_n$ and $\PLeg_l\Delta\mu$ 
do not depend on $T$ and are completely determined 
by the magnetic field configuration, $\vec{B}(r,\, \theta)$. 
In contrast, the zeroth Legendre components $\PLeg_0\delta\mu_n$ and 
$\PLeg_0\Delta\mu$ cannot be determined in this way 
since they depend on an arbitrary function $\zeta(r)$ in Eqs.~(\ref{eq:naAlgProb}). 
The latter function can be found from the requirement that $U_{b\theta} = 0$ 
on the symmetry axis for arbitrary $r$ \cite{GKO17}, i.e.,
\begin{align}
U_{b\theta} = 0 \quad {\rm at } \quad \theta=0 \quad  {\rm and} \quad \theta = \pi.
\label{req1}
\end{align}
Expanding the function $V(r,\, \theta)$ in Legendre polynomials, 
$V = \sum_{l=0}^\infty (\PLeg_l V) P_l(\cos\theta)$, 
and using Eqs.~(\ref{eq:Vdef}) and (\ref{eq:Upfin}), one has
\begin{equation}
\label{eq:Utheta-Leg}
U_{b\theta} = -\frac{1-\cos\theta}{r n_b \sin\theta}\left(  n_b r^2 \PLeg_0 V \right)' + \frac{1}{r n_b}\sum_{l=1}^\infty \left( n_b r^2 \PLeg_l V \right)' \frac{P_l^1(\cos\theta)}{l(l+1)},
\end{equation}
where $P_l^1(\cos\theta)$ is the associated Legendre polynomial.
Since $V$ and $V'$ should be finite everywhere in the core, 
we see that (\ref{req1}) is automatically satisfied for $\theta = 0$, 
while for $\theta=\pi$ it requires 
\begin{equation}
\label{eq:P0Vtmp}
\left( n_b r^2 \PLeg_0 V \right)' = 0 \Longleftrightarrow n_b r^2 \PLeg_0 V = C
\end{equation}
where $C$ is a constant. 
Taking this expression at $r = 0$ 
and recalling again that $V$ is finite, we obtain $C=0$. 
Therefore, $\PLeg_0 V = 0$ for any $r$. 
Note that we could arrive at the same result by considering 
the baryon conservation law (\ref{eq:nbcont}) in the integral form.
Taking into account that $U_{br}=V$ (see Eq.\ \ref{eq:Upfin}),
this condition simply means that $\PLeg_0 U_{br} = 0$.
In fact, it is not difficult to prove that the same condition is also true for a 
net flow velocity ${\pmb U}$, introduced in Sec.\ \ref{sec:sol},    
\begin{align}
\PLeg_0 U_{br} = 0, \quad \quad \PLeg_0 U_{r} = 0.
\label{res1}
\end{align}
Looking 
at the definition (\ref{eq:Vdef}) of the function $V$, 
we see that the condition $\PLeg_0 V = 0$ 
depends on both the functions $\PLeg_0\delta\mu_n$ and $\PLeg_0\Delta\mu$. 
Hence, we need an additional equation 
in order to close the system (to express $\PLeg_0\delta\mu_n$ through $\PLeg_0\Delta\mu$). 
This additional equation is provided by the 
zeroth Legendre component of Eq.~(\ref{eq:fAbalance}). 
As a result, we have two differential equations for two unknowns, 
$\PLeg_0\delta\mu_n$ and $\PLeg_0\Delta\mu$,
which can be written as
\begin{subequations}
\label{eq:P0muFin}
\begin{align}
&\frac{1}{r^2}\frac{\diff}{\diff r}\left[ r^2 \frac{n_e n_n^2}{n_b J_{np}} \left( \PLeg_0 \delta\mu_n \right)' \right] + \PLeg_0\Delta\Gamma(\Delta\mu) = 0, \label{eq:P0muFin-long}\\
&n_b \left( \PLeg_0\delta\mu_n \right)' + n_e \left( \PLeg_0\Delta\mu \right)' = \PLeg_0 f_{Ar}, \label{eq:P0muFin-short}
\end{align}
\end{subequations}
where 
we make use of the equalities
$\partial n_a/\partial\theta=0$ 
and $\partial J_{np}/\partial\theta=0$.
Eq.~(\ref{eq:P0muFin-short}) here
is a simple 1st-order differential equation,
while 
Eq.~(\ref{eq:P0muFin-long})
is of the
2nd-order, with variable coefficients and, in general, 
nonlinear dependence of $\Delta\Gamma$ on $\Delta\mu$.
Thus, this system hardly has an analytic solution 
and should generally be solved numerically.
But first, it should be supplied with a number of boundary conditions,
which are discussed in the next section.
 
Note that the method of calculation of the quantities 
$\PLeg_0\delta\mu_n$ and $\PLeg_0\Delta\mu$, 
suggested here, 
is an (improved and simplified) version of the general method 
presented in appendix~D of Ref.\ \cite{GKO17}. 
In particular, the left-hand side of Eq.~(\ref{eq:P0Vtmp}) 
is equivalent to equation (D10) in \cite{GKO17}, 
Eq.~(\ref{eq:P0muFin-short}) coincides with equation (D13), 
and Eq.~(\ref{eq:P0muFin-long}) is the equation~(D11) 
integrated with the boundary condition (D12).

\subsection{Boundary conditions for Eqs.~(\ref{eq:P0muFin})}
\label{sec:sol-bound}

The system of Eqs.~(\ref{eq:P0muFin}) requires three boundary conditions. 
Two of them should be chosen at the stellar center,
\begin{equation}
\label{eq:bound-cent}
\left( \PLeg_0\delta\mu_n \right)'\Bigr|_{r \rightarrow 0} = 0, 
\qquad \PLeg_0\delta\mu_n\bigr|_{r \rightarrow 0} = C_1.
\end{equation}
The first of these conditions follows from the regularity requirement of
$\PLeg_0\Delta\Gamma$ at 
$r\rightarrow 0$
(see Eq.~\ref{eq:P0muFin-long}). 
Due to axisymmetry of the problem we have $\PLeg_l f_{Ar}|_{r \rightarrow 0} = 0$, 
thus, in view of Eq.\ (\ref{eq:P0muFin-short}), this condition is equivalent 
to $(\PLeg_0 \Delta\mu)'|_{r \rightarrow 0} = 0$ 
(see also equation (D14) in Ref.\ \cite{GKO17}). 
The second condition 
determines the value of the perturbed neutron chemical potential at the centre.%
%
\footnote{
Indeed, one can see from Eq.~(\ref{eq:DmudmuHighL}) that $\PLeg_l\delta\mu_n|_{r \rightarrow 0} = 0$ 
for $l\geqslant 1$, hence
$\PLeg_0\delta\mu_n|_{r \rightarrow 0} = \delta\mu_n|_{r \rightarrow 0}$. 
}
%
The constant $C_1$ there specifies the total 
number of baryons (or the central density) in the perturbed star. 
In principle, if one studies the magnetic field evolution in time, 
one should adjust $C_1$ at each time step in order to conserve the total 
number of baryons in the star.
However, here we consider  an NS at some particular moment of its evolution.
For our purposes, therefore, it is 
sufficient to set, for example, $C_1=0$.

To obtain the third (and the last) boundary condition one should
match the solution of Eqs.~(\ref{eq:P0muFin}) with 
the corresponding solution of similar equations in the crust \cite{GKO17}.  
Such an analysis is beyond the scope of the present paper.
Instead, here we employ
a simplified method pointed out in Ref.~\cite{GKO17}.
It allows us to formulate 
an approximate expression for
the required boundary condition.
To proceed, let us assume that 
the quasistationary approximation is valid not only in the core
but also in the NS crust.
Then, using integral form of the continuity equation (\ref{eq:MHDquas-cont}) 
for, e.g., neutrons, one obtains
\begin{equation}
\label{eq:cc-nSave}
\int_\text{core} \Delta\Gamma \diff V + \int_\text{crust} \Delta\Gamma_{n\,\text{crust}} \diff V = 0,
\end{equation}
where $\Delta\Gamma_{n\,\text{crust}}$  represents schematically
the total rate of reactions with free neutrons in the crust. 
To rigorously calculate this quantity,
one has to formulate a system of quasistationary evolution equations in the crust and solve it.
However, one can avoid this complication by assuming
that the neutron reactions in the crust 
are much less efficient than in the core,
so that the second term in Eq.~(\ref{eq:cc-nSave}) can be neglected. 
Then the third boundary condition will take the form
(cf. the end of appendix D in Ref.\ \cite{GKO17})
\begin{subequations}
\label{eq:cc-Slow}
\begin{equation}
\label{eq:cc-SlowInt}
\int_0^{R_\text{core}} \left( \PLeg_0 \Delta\Gamma \right) r^2 \diff r = 0,
\end{equation}
where $R_\text{core}$ is the core radius.
To derive this formula 
we expanded $\Delta \Gamma$ in Legendre polynomials and integrated it over $\theta$.
Integrating then Eq.~(\ref{eq:P0muFin-long}) and using (\ref{eq:cc-SlowInt}),
we can represent 
it
in the equivalent differential form,
\begin{equation}
\label{eq:cc-SlowDiff}
\left( \PLeg_0 \delta\mu_n \right)'\Bigr|_{R_\text{core}} = 0.
\end{equation}
\end{subequations}
Eqs.~(\ref{eq:bound-cent}) and (\ref{eq:cc-SlowDiff}) constitute 
a complete set of boundary conditions for Eqs.~(\ref{eq:P0muFin}). 

Eqs.~(\ref{eq:P0muFin}) can be solved analytically 
in two limiting cases. Before discussing them, 
let us introduce the coefficient $\lambda = \Delta\Gamma/\Delta\mu$.
According to Refs.\ \cite{Reis95,ykgh01}, 
$\lambda$ does not depend on $\Delta\mu$ 
if $\Delta\mu\ll \pi kT$, where $k$ is the Boltzmann constant. 
As argued in Ref.~\cite{GKO17}, and follows 
from the analysis of the system (\ref{eq:P0muFin}),
its solution is governed by the
only one dimensionless parameter, 
$\varkappa = n_e^2 n_n^2/(n_b^2 J_{np}L_B^2\lambda)$.
The friction coefficient scales as $J_{np}\propto \Tg^2$ \cite{ys90},
while $\lambda \propto \Tg^6$ for the modified Urca processes
($\Tg$ is the redshifted temperature, see Sec.~\ref{sec:figures} for more details). Thus, $\varkappa\propto \Tg^{-8}$. 
When $\Tg$ is low enough, i.e., $\varkappa\gg 1$, 
the last term in Eq.~(\ref{eq:P0muFin-long}) can be neglected
and, with the boundary conditions (\ref{eq:bound-cent}), 
we have $\PLeg_0\delta\mu_n = 0$. 
Then, using Eq.~(\ref{eq:P0muFin-short}) and the boundary condition (\ref{eq:cc-SlowInt}), 
we can find $\PLeg_0\Delta\mu$. 
In the opposite limit of high enough $\Tg$, 
i.e., $\varkappa \ll 1$, 
the first term in Eq.~(\ref{eq:P0muFin-long}) 
is negligible in comparison to the last one, 
hence $\PLeg_0\Delta\Gamma = 0$. 
The high temperature
means that $\Delta\mu\ll \pi kT$,
so that the condition $\PLeg_0\Delta\Gamma = 0$ is equivalent to $\PLeg_0\Delta\mu = 0$.
The solution for $\PLeg_0\delta \mu_n$ can then be 
found by integrating Eq.~(\ref{eq:P0muFin-short}).
Note, however, that 
$\PLeg_0\delta \mu_n$, obtained in this way, 
is incompatible with the boundary condition (\ref{eq:cc-SlowDiff}).
Accurate solution to Eqs.~(\ref{eq:P0muFin})
would give a slightly different 
$\PLeg_0\delta \mu_n$
in the very vicinity
of the crust-core interface, 
compatible with Eq.~(\ref{eq:cc-SlowDiff});
see Sec.~\ref{sec:figures-mumu} for more details.

In this section we formulated a scheme that allows us
to calculate all the Legendre components of $\delta\mu_n$ and $\Delta\mu$. 
These quantities should then be substituted into Eq.~(\ref{eq:EcomSol}) 
in order to determine $\delta\vec{E}_\text{com}$
and into Eqs.~(\ref{eq:Vdef}) and (\ref{eq:Upfin}) to determine $U_{br}$ and $U_{b\theta}$. 
The last quantity, that remains to be found, 
is the toroidal component of the baryon velocity,~$U_{b\varphi}$.

\subsection{Toroidal component of the baryon velocity and evolution of the magnetic field $\vec{B}$}
\label{sec:sol-Uphi}

The method to calculate the toroidal component of $\vec{U}_b$,
proposed here, 
is equivalent to the one sketched in section~III$\,$C$\,$2 of Ref.\ \cite{GKO17}.
Here we discuss it in more detail, 
using the well known representation of the axisymmetric magnetic field 
through the poloidal flux and poloidal current functions \cite{Goed2010}.

Let us present the magnetic field as a sum of
poloidal ``$\pind$'' and toroidal ``$\tind$'' components,
$\vec{B} = \vec{B}^\pind + \vec{B}^\tind$.
In the axisymmetric case one can introduce the scalar functions 
$\Psi(r,\theta)$ and $I(r,\theta)$ such that 
(see, e.g., Ref.~\cite{Goed2010}; cf. Sec.~\ref{sec:sol-Up})
\begin{equation}
\label{eq:BpBt}
\vec{B}^\pind = \vec{\nabla}\Psi\times\vec{\nabla}\varphi = \rot\left( \Psi\vec{\nabla}\varphi \right), \qquad \vec{B}^\tind = I\vec{\nabla}\varphi,
\end{equation}
where $\Psi(r,\,\theta)$ is the poloidal flux function. 
Its name is due to the fact that 
$2\pi\Psi(r,\theta)$ is the magnetic flux passing throw the polar cap 
with radius $r$ and opening angle $\theta$. 
The curves $\Psi ={\rm const}$ are the poloidal magnetic field lines. 
The vector potential for $\vec{B}^\pind$ is $\Psi\vec{\nabla}\varphi$. 
In turn,  $I(r,\,\theta)$ is the poloidal current function, 
since $cI(r,\theta)/2$ 
is the electric current passing throw the same polar cap. 
The axisymmetric magnetic field is fully determined by specifying the functions  
$\Psi(r,\theta)$ and $I(r,\theta)$. 

Note that in the consideration above we only used 
the poloidal part of the force balance Eq.~(\ref{eq:fAbalance}). 
For the toroidal part we have (see Appendix~\ref{app:dtPsi} for details)
\begin{equation}
\label{eq:fAtor}
\vec{f}_A^\tind = -\frac{\left(\vec{\nabla}\varphi\right)^2}{4\pi}\vec{\nabla}\Psi\times\vec{\nabla}I = 0 \quad \Longrightarrow \quad I = I(\Psi,t),
\end{equation}
i.e., $I$ depends on the coordinates $r$ and $\theta$ only through 
the poloidal flux function $\Psi(r, \, \theta,\, t)$ \cite{Goed2010}.%
%
\footnote{As a consequence, 
if the magnetic field is force-free in the toroidal direction, 
its poloidal and toroidal components are rigidly coupled: 
$\vec{B}^\tind$ is constant on the poloidal field lines.}
%
This means, in particular, that
\begin{subequations}
\label{eq:Ideriv}
\begin{align}
\vnabla I(r,\, \theta,\, t) &= I'_\Psi (\Psi,t) \, \vnabla\Psi(r,\, \theta,\, t), \label{eq:Ideriv-r}\\
\qquad \pd{I(r,\, \theta,\, t)}{t} &= I'_\Psi (\Psi,t) \, \pd{\Psi(r,\, \theta,\, t)}{t} + I'_t (\Psi,t), \label{eq:Ideriv-t}
\end{align}
\end{subequations}
where $I'_\Psi(\Psi,\,t)$ is the partial derivative of $I(\Psi,\,t)$ with respect to $\Psi$ at constant $t$
and $I'_t(\Psi,\,t)$ is the partial derivative of $I(\Psi,\,t)$ with respect to $t$ at constant $\Psi$. 
Note that, similar to the function $I(\Psi,\, t)$,
both the functions $I'_\Psi(\Psi,\, t)$ and $I'_t(\Psi,\, t)$
depend on the spatial coordinates $r$ and $\theta$ 
only through the function $\Psi(r,\,\theta)$.
This very important property, that will be used in what follows,
is equivalent to the requirement that not only $\vec{f}_A^\tind$,
but also $\partial \vec{f}_A^\tind/\partial t$ must vanish 
in an axisymmetric quasistationary NS \cite{GKO17}.

As a first step towards calculation of $\vec{U}_{b}^\tind$, let us decompose Eq.~(\ref{eq:faradCom}) into poloidal and toroidal parts
\begin{subequations}
\label{eq:faradSplit}
\begin{align}
\pd{\vec{B}^\pind}{t} &= -c\rot\delta\vec{E}_\text{com}^\tind + \rot\left( \vec{U}_b^\pind\times\vec{B}^\pind \right), \label{eq:faradSplit-pol}\\
\pd{\vec{B}^\tind}{t} &= -c\rot\delta\vec{E}_\text{com}^\pind + \rot\left( \vec{U}_b^\pind\times\vec{B}^\tind \right) + \rot\left( \vec{U}_b^\tind\times\vec{B}^\pind \right). \label{eq:faradSplit-tor}
\end{align}
\end{subequations}

Equation (\ref{eq:faradSplit-pol}) here contains the quantities $\delta\vec{E}_\text{com}$ and $\vec{U}_b^\pind$,
which are already calculated
(see Eqs.~\ref{eq:EcomSol} and \ref{eq:Vdef} supplemented with 
expressions for $\delta\mu_n$ and $\Delta\mu$ from Secs.~\ref{sec:sol-mu} and \ref{sec:sol-bound}). 
In terms of the flux function $\Psi$, it can be rewritten as
\begin{equation}
\label{eq:PsiEvol}
\pd{\Psi}{t} = -\left( \vec{U}_b^\pind + \frac{n_n^2}{n_b J_{np}}\vec{\nabla}\delta\mu_n \right)\cdot\vec{\nabla}\Psi + \frac{c^2 J_{ep}}{4\pi e^2 n_e^2}\grsh\Psi,
\end{equation}
where $\grsh$ is the Grad-Shafranov operator (see Appendix~\ref{app:dtPsi} for details). 
The first term in brackets describes advection of the field lines by the fluid motions; 
it also includes the effects of the magnetic field dissipation due to nonequilibrium beta-processes.
The second term describes evolution due to the ambipolar diffusion. 
Finally, the last term in Eq.~(\ref{eq:PsiEvol}) is responsible  for Ohmic dissipation. 
Generally, Eq.\ (\ref{eq:PsiEvol}) is similar to equation~(7) of Ref.\ \cite{GourCum14}, 
which is discussed in the context of the magnetic field evolution in the NS crust. 
The only difference is that, instead of $\vec{U}_b^\pind$,
equation~(7) features the electron poloidal velocity,
while the term proportional to ${\pmb \nabla} \delta \mu_n$ is absent.
If the field is frozen-in to the fluid, 
i.e. $\delta\vec{E}_\text{com} = 0$, then 
only the term $-\vec{U}_{b}^\pind \cdot \vnabla\Psi$ 
in Eq.~(\ref{eq:PsiEvol}) 
survives,
so that Eq.~(\ref{eq:PsiEvol})
transforms into the form 
similar to that used in Ref.\ \cite{blb2017} 
to model the magnetic flux expulsion from the superconducting NS core.

Now let us consider Eq.~(\ref{eq:faradSplit-tor}) for the toroidal magnetic field. 
Using Eqs.\ (\ref{eq:BpBt}) and (\ref{eq:Ideriv-t}), it can be rewritten as  
\begin{equation}
\label{eq:dtIDphi}
\left( I'_\Psi\pd{\Psi}{t} + I'_t \right)\vnabla\varphi = \rot\left( -c\delta\vec{E}_\text{com}^\pind + I \, \vec{U}_b^\pind\times\vnabla\varphi + \vec{U}_b^\tind\times\vec{B}^\pind \right),
\end{equation}
After some manipulations (see Appendix~\ref{app:Uphi}), 
Eq.~(\ref{eq:dtIDphi}) takes the form
\begin{equation}
\label{eq:It}
(\vnabla\varphi)^2 I'_t = F(r,\theta) + \vec{B}^\pind\cdot\vnabla\left( G(r,\theta) + \vnabla\varphi\cdot\vec{U}_b^\tind \right),
\end{equation}
where
\begin{subequations}
\label{eq:FandG}
\begin{align}
F &= (\vnabla\varphi)^2 \vnabla\Psi\cdot\vnabla\left( \frac{c^2 J_{ep}}{4\pi e^2 n_e^2} I'_\Psi \right) - I\diver\left[ (\vnabla\varphi)^2\left( \frac{n_n^2}{n_b J_{np}}\vnabla\delta\mu_n + \vec{U}_b^\pind \right) \right], \label{eq:FandG-F}\\
G &= \frac{c (\vnabla\varphi)^2}{4\pi e n_e}\left( \grsh\Psi + I I'_\Psi \right). \label{eq:FandG-G}
\end{align}
\end{subequations}
It can be further rewritten as
\begin{align}
\vec{B}^\pind\cdot\vnabla\left(\frac{U_{b \varphi}}{r \, {\rm sin}\theta} \right)=S(r,\, \theta),
\label{Ubphi}
\end{align}
or as
\begin{align}
B^\pind \frac{\partial }{\partial \chi_\Psi}
\left(\frac{U_{b \varphi}}{r \, {\rm sin}\theta} \right)=S(r,\, \theta),
\label{Ubphi2}
\end{align}
where $\chi_\Psi$ is the ``magnetic'' coordinate (length) 
along the poloidal field line $\Psi(r,\,\theta) ={\rm const}$ \cite{Goed2010}
(we remind the reader that $\vec{B}^\pind$ is directed along these lines);
\begin{align}
S(r,\,\theta) = (\vnabla\varphi)^2 I'_t -F(r,\theta)- \vec{B}^\pind\cdot\vnabla G(r,\theta),
\label{S}
\end{align}
and we make use of the identity 
$\vnabla\varphi \cdot \vec{U}_b^\tind = U_{b\varphi}/(r\sin\theta)$.
Equation (\ref{Ubphi2}) can be easily integrated,
the result is
\begin{equation}
\label{eq:UphiGen}
U_{b\varphi} = r\sin\theta \left( - G\bigr|_{\chi_{0\Psi}}^{\chi_{\Psi}} - \int_{\chi_{0\Psi}}^{\chi_\Psi} \frac{F}{B^\pind}\diff\chi + I'_t(\Psi,t) \int_{\chi_{0\Psi}}^{\chi_{\Psi}}\frac{(\vnabla\varphi)^2}{B^\pind}\diff\chi + \Omega_0(\Psi,t) \right),
\end{equation}
where $\Omega_0(\Psi,t) = U_\varphi/(r\sin\theta)|_{\chi_{0\Psi}}$ 
is the boundary condition at the point $\chi_{0\Psi}$. 
Integration in Eq.\ (\ref{eq:UphiGen}) 
is performed from a starting point $\chi_{0\Psi}$ on the line 
with a given $\Psi$ up to some point $\chi_{\Psi}$ on this line.
Note that, to obtain Eq.~(\ref{eq:UphiGen}) we make use of the fact 
that $I'_t(\Psi,t)$ is constant on the poloidal field lines 
(and hence can be taken outside the corresponding integral, see the third term 
in the brackets in Eq.\ \ref{eq:UphiGen}).
In contrast to the functions $I(\Psi,t)$ and $I'_\Psi(\Psi,t)$,
whose dependence on the coordinates $r$  and $\theta$ 
is completely specified by the instantaneous configuration of the magnetic field, the function $I'_t(\Psi,t)$ is generally not known 
(depends on $\partial \vec{B}/\partial t$).
Hence, generally, $U_{b\varphi}$ is determined on each field line up to two
``constants'', $I'_t(\Psi,t)$ and $\Omega_0(\Psi,t)$. However, $I'_t(\Psi,t)$
can be found from the boundary conditions, as explained below.

\begin{figure}
\includegraphics[width=0.45\textwidth]{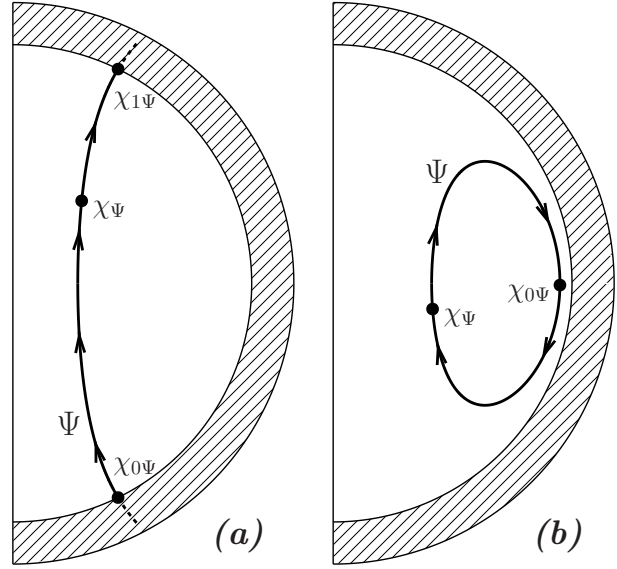}
\caption{\label{fig:lines}%
Meridional cross-sections of an NS. 
Straight vertical lines show the symmetry axis of the star. 
Hatched regions show the NS crust, white regions show the core. 
The poloidal magnetic field lines are depicted by the
thick black lines (lie on the surface of constant $\Psi$). 
A direction of the field is shown by arrows and coincides 
with the direction of integration in Eqs.~(\ref{eq:UphiGen}), (\ref{eq:ItFin}), and (\ref{eq:ItFin2}). 
\textit{(a)} The field line with a given $\Psi$ is open in the core. 
The initial point $\chi_{0\Psi}$ coincides with the point, 
where the field line enters the core. 
A point where the field line leaves the core is denoted as $\chi_{1\Psi}$. 
\textit{(b)} The field line with a given $\Psi$ is closed in the core. 
The starting point $\chi_{0\Psi}$ is defined as the nearest point to the crust. 
}
\end{figure}

There are two possibilities. 
The first one is that the field line is open in the core 
(Fig.~\ref{fig:lines}a) and has two points where it crosses the crust-core interface. 
Here we do not care whether this line closes up in the crust or continues to the magnetosphere.%
%
\footnote{The magnetosphere field near the NS surface should resemble a vacuum one. 
The reason for that is very low charge current density in the magnetosphere 
(e.g., Ref.~\cite{BGI2006}). 
As a consequence, $I=0$ on the field lines that 
do not close up inside the star 
(since the condition \ref{eq:fAtor} 
should be valid both in the core, crust and 
close
magnetosphere). 
This observation allows one to simplify Eqs.~(\ref{eq:UphiGen}) and (\ref{eq:ItFin}) considerably.}
%
Let $\chi_{0\Psi}$ be a point where the field line enters the core from the crust, 
and $\chi_{1\Psi}$ be a point where the field line leaves the core. 
There should be a boundary condition at $\chi_{1\Psi}$ 
similar to that at the point $\chi_{0\Psi}$, namely,  
$U_{b\varphi}/(r\sin\theta)|_{\chi_{1\Psi}} = \Omega_1(\Psi,t)$. 
Then we have for $I'_t(\Psi,\, t)$ (see Eq.\ \ref{eq:UphiGen})
\begin{subequations}
\label{eq:ItFin}
\begin{equation}
\label{eq:ItFin-Open}
I'_t(\Psi,t) = \left( \int_{\chi_{0\Psi}}^{\chi_{1\Psi}} \frac{(\vec{\nabla}\varphi)^2}{B^\pind}\diff\chi \right)^{-1} \left( \Omega_1(\Psi,t) - \Omega_0(\Psi,t) + G\bigr|_{\chi_{0\Psi}}^{\chi_{1\Psi}} + \int_{\chi_{0\Psi}}^{\chi_{1\Psi}} \frac{F}{B^\pind}\diff\chi \right).
\end{equation}

The second option is realized if the field line is closed up in the core (Fig.~\ref{fig:lines}b). 
Integrating then Eq.~(\ref{eq:UphiGen}) over the closed loop along the field line
shown in Fig.~\ref{fig:lines}b, we obtain
\begin{equation}
\label{eq:ItFin-Close}
I'_t(\Psi,t) = \left( \oint_\Psi \frac{(\vec{\nabla}\varphi)^2}{B^\pind}\diff\chi \right)^{-1} \oint_\Psi \frac{F}{B^\pind}\diff\chi.
\end{equation}
\end{subequations}
Note that there is no way to relate the ``boundary'' function $\Omega_0(\Psi,t)$ 
to that in the crust in the case of closed field lines.
The region in the core, where the field lines are closed up, 
appears to be magnetically decoupled from the rest of the star. 
This seems to be in line with the results of Ref.\ \cite{gl15}, 
although the arguments presented there are quite different.

Note also that for a magnetic field configuration with equatorial symmetry
one has $\Omega_1(\Psi,\, t)=\Omega_0(\Psi,\, t)$ and $G|_{\chi_{1\Psi}}=G|_{\chi_{0\Psi}}$,
and consequently $I'_t(\Psi,t)$ is independent of the ``boundary'' functions
$\Omega_0(\Psi,\, t)$ and $\Omega_1(\Psi,\, t)$ (cf.\ Eq.~\ref{eq:ItFin-Open}),
\begin{equation}
\label{eq:ItFin2}
I'_t(\Psi,t) = \left( \int_{\chi_{0\Psi}}^{\chi_{1\Psi}} \frac{(\vec{\nabla}\varphi)^2}{B^\pind}\diff\chi \right)^{-1} \int_{\chi_{0\Psi}}^{\chi_{1\Psi}} \frac{F}{B^\pind}\diff\chi.
\end{equation}

Summarizing, we have a recipe to derive the function $I'_t(\Psi,\, t)$ in Eq.~(\ref{eq:UphiGen}). 
Therefore, we can calculate $U_{b\varphi}$ for a given magnetic field configuration.

\subsection{Summary of results for Sec.~\ref{sec:sol} }
\label{sec:sol-concl}

To calculate the fluid motions induced by the magnetic field one should:

(i) determine $\Delta\mu$ and $\delta\mu_n$ 
from Eqs.~(\ref{eq:DmudmuHighL}) and (\ref{eq:P0muFin}) 
with the boundary conditions (\ref{eq:bound-cent})
and (\ref{eq:cc-Slow});

(ii) calculate $\vec{U}_b^\pind$  from Eqs.~(\ref{eq:Vdef}) and (\ref{eq:Upfin}), 
and $\vec{U}_b^\tind$ from Eqs.~(\ref{eq:UphiGen}) and (\ref{eq:ItFin});

(iii) if one also needs the particle species velocities $\vec{u}_a$
one should first calculate the diffusion velocities $\vec{w}_a$ from Eqs.~(\ref{eq:Wsol}) 
and then use the formula $\vec{u}_a =\vec{U}_b+\vec{w}_a$.

Moreover, from Eqs.~(\ref{eq:PsiEvol}) and (\ref{eq:ItFin}) 
we now know the time derivatives $\partial\Psi/\partial t$ and $I'_t(\Psi,\, t)$, 
which opens up a possibility  
to determine the derivative $\partial\vec{B}/\partial t$ 
for arbitrary  $\vec{B}$, 
and hence to model the magnetic field evolution 
in a fully self-consistent way.

\section{Quasistationary particle flows: numerical results}
\label{sec:figures}

The general formalism developed in the previous section allows 
one to calculate particle velocities in a nonsuperfluid magnetized $npe$-cores of NSs. 
In this section we present an example of such calculation 
for two simple models of the poloidal magnetic field, and discuss the main properties 
of the obtained solutions.

We adopt the same microphysical inputs  
as in our previous work \cite{GKO17}. 
Namely, we consider an NS with the mass  
$M=1.4\,$M$_\odot$ and assume $npe$ HHJ~equation of state in the core from Ref.\ \cite{hhj99}. 
Such an NS model has a radius $R=12.2$~km and the core radius $R_{\rm core}=11.2$~km.
To calculate the unperturbed quantities as functions of the radial coordinate $r$ 
[e.g., the number densities, $n_a(r)$]
we,
somewhat inconsistently, 
use fully relativistic Tolman-Oppenheimer-Volkoff equations \cite{ov39,t39}, 
although perturbed equations are treated in the Newtonian framework.
Such an approach
allows us to obtain reasonable NS radii,  
while keeping the discussion of the magnetic field effects relatively simple.
We do not expect that accurate account for 
General Relativity effects will affect our results qualitatively.

To calculate the friction coefficients, $J_{ep}$ and $J_{np}$, 
we employ equations (78) and (79) from Ref.\ \cite{GKO17}.
In turn, the reaction rate $\Delta\Gamma$, 
which is generated by the modified Urca processes 
is taken from Ref.\ \cite{ykgh01}%
%
\footnote{ 
Note that the angular integral for a proton branch of modified Urca process 
in Ref.\ \cite{ykgh01} should be corrected as it is described, e.g., in Ref.~\cite{kyh16}.}
%
(the more powerful direct Urca process is forbidden for the chosen NS configuration).
All these quantities 
depend on the local stellar temperature $T$,
which is related to the redshifted stellar temperature $\Tg$
by the formula: $T=\Tg/\sqrt{g_{00}}$, 
where $g_{00}(r)$ is the time component of the metric tensor.
Because of high thermal conductivity (see, e.g., Ref.~\cite{ppp15}), 
the cores of not too young NSs can be treated 
as nearly isothermal, $\Tg=\const$, 
which makes $\Tg$ a convenient parameter 
to characterize the thermal state of our system (see below).

\subsection{``Minimal'' model of the magnetic field}
\label{sec:figures-minField}

\begin{figure}
\begin{minipage}[t]{0.49\textwidth}
\includegraphics[height=0.35\textheight]{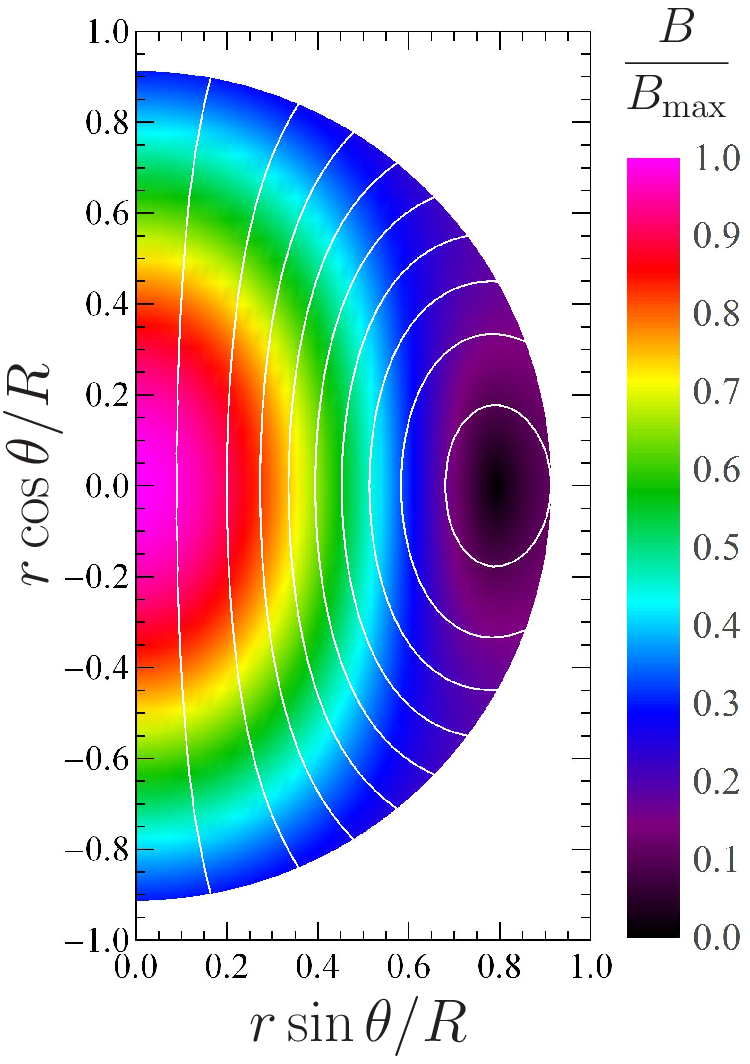}
\caption{\label{fig:B-min} A meridional cross-section of an NS core
and ``minimal'' configuration of the magnetic field considered in the text. 
Field lines are shown by white curves. 
The strength of the magnetic field is denoted by different colors.}
\end{minipage}
\hfill
\begin{minipage}[t]{0.49\textwidth}
\includegraphics[height=0.35\textheight]{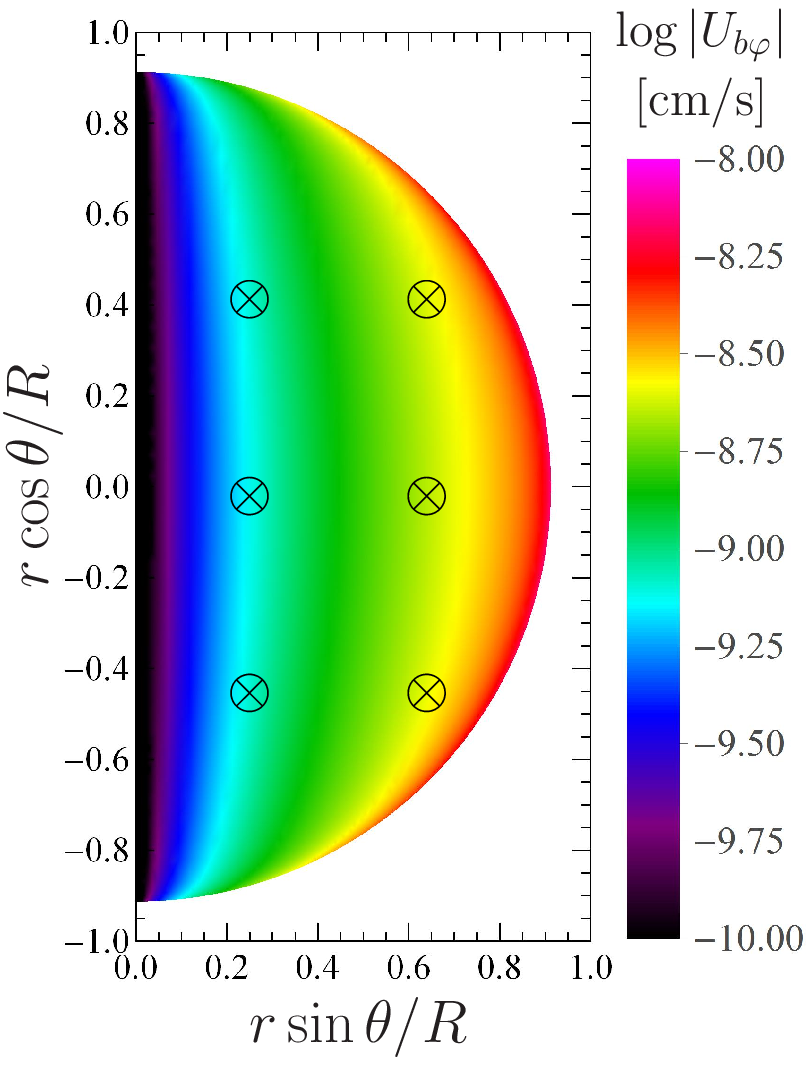}
\caption{\label{fig:Uphi-min} 
The toroidal component 
of the baryon velocity $U_{b \varphi}$ for the minimal magnetic field model. 
Colors denote the value of $U_{b \varphi}$ in the logarithmic scale (in cm/s).
Crossed circles represent the direction of $U_{b \varphi}$ (cf. Fig.\ \ref{fig:fBUphi-noFlow}c). 
}
\end{minipage}
\end{figure}

Following Refs.\ \cite{armm13,papm17} we start with the simplest analytic expression
fo the poloidal flux function,
\begin{equation}
\label{eq:PsiThis}
\Psi(r,\theta) = B_\text{max} R^2 f\left( \frac{r}{R} \right)\sin^2\theta,
\end{equation}
where $B_\text{max}$ is the maximum value of the magnetic field inside the star;
$R$ is the NS radius;
and $f(x)$ is the analytic function to be specified below.
In the axisymmetric case the field near the stellar centre 
should be almost homogeneous and aligned with the symmetry axis. 
Thus at $x=r/R \ll 1$ we should have $f\sim x^{2n}$, where $n\geq 1$ 
(the case $n=0$ is excluded
by the regularity requirement of the magnetic field at $x \rightarrow 0$).
Assuming, as was suggested in Ref.\ \cite{armm13},
that $f$ inside the star takes the form, $f=\sum_{n\geq 1} f_{2n}\, x^{2n}$,
and matching this expression with the vacuum dipole field outside the star ($f \propto 1/x$),
we obtain \cite{papm17}
\begin{equation}
\label{eq:f-min}
f(x) = \frac{1}{2}x^2 - \frac{3}{5}x^4 + \frac{3}{14}x^6.
\end{equation}
This is a minimal solution for $f(x)$ in the domain $x\in[0,1]$,
satisfying the requirements
\begin{equation}
\label{eq:fBounds-surf}
f'(1) = -f(1), \qquad f''(1) = 2f(1).
\end{equation}
The first of these conditions is necessary for 
continuity of the magnetic field,
while the second one ensures that the charge current density at the surface vanishes
(see, e.g., Ref.\ \cite{papm17} for more details).
The expression (\ref{eq:f-min}) differs from that proposed in equation (30) of Ref.\ \cite{papm17}
by a numerical factor $35/4$, appearing due to a different normalization.
The magnetic field $\vec{B} = \vec{B}^\pind$,
corresponding to $f(x)$ from Eq.\ (\ref{eq:f-min}),
is shown in Fig.~\ref{fig:B-min}. 
%

Now let us turn to the toroidal (azimuthal) velocity $U_{b\varphi}$.
It is interesting that $U_{b\varphi}$ can be easily calculated 
for a purely poloidal field 
(when the poloidal current function $I=0$). 
Then, according to Eqs.~(\ref{eq:FandG}) and (\ref{eq:app:jPsi}), 
one has $F=0$ and 
$G = - j_\varphi / (e n_e r \sin\theta)$;
the function $I'_t(\Psi,\, t)$ also vanishes because of the equatorial symmetry
of the chosen magnetic field (see Eq.\ \ref{eq:ItFin2}).
Hence, $U_{b\varphi}$ can be found provided that the ``boundary'' function 
$\Omega_0(\Psi,\, t)$ is known (see Eq.\ \ref{eq:UphiGen}).
Generally, $\Omega_0(\Psi,\, t)$ can only be determined 
from the joint solution to the full system
of the magnetic evolution equations in the crust and core.
Here we do not attempt to find such a solution 
and instead, for illustration, assume
that $\Omega_0(\Psi) = -G|_{\chi_{0\Psi}}$. Then
\begin{equation}
\label{eq:Uphi-pol}
U_{b\varphi} = \frac{j_\varphi}{e n_e}=- \frac{c}{4\pi e n_e \, r {\rm sin}\,\theta} \grsh\Psi;
\end{equation}
the result is plotted in Fig.~\ref{fig:Uphi-min}.
Note that $U_{b\varphi}$ does not explicitly depend on the NS temperature 
and is proportional to $B_{\rm max}$. Note also that, 
with our choice of boundary condition for $U_{b\varphi}$,
it follows from Eqs.\ (\ref{eq:Wsol-e}) and (\ref{eq:Uphi-pol}) that 
$U_{b \varphi}$ is related to the $\varphi$-component of 
the electron diffusion velocity $\vec{w}_e$ by the formula,
\begin{align}
U_{b \varphi}=-w_{e\varphi}.
\label{wephi}
\end{align}
%

\subsection{Chemical potential perturbations}
\label{sec:figures-mumu}

Now let us discuss how the application of the method,
outlined in Secs.~\ref{sec:sol-mu} and \ref{sec:sol-bound},  
works for calculation of the chemical potential perturbations, 
$\delta\mu_n$ and $\Delta\mu$. 

\begin{figure}
	\includegraphics[width=\textwidth]{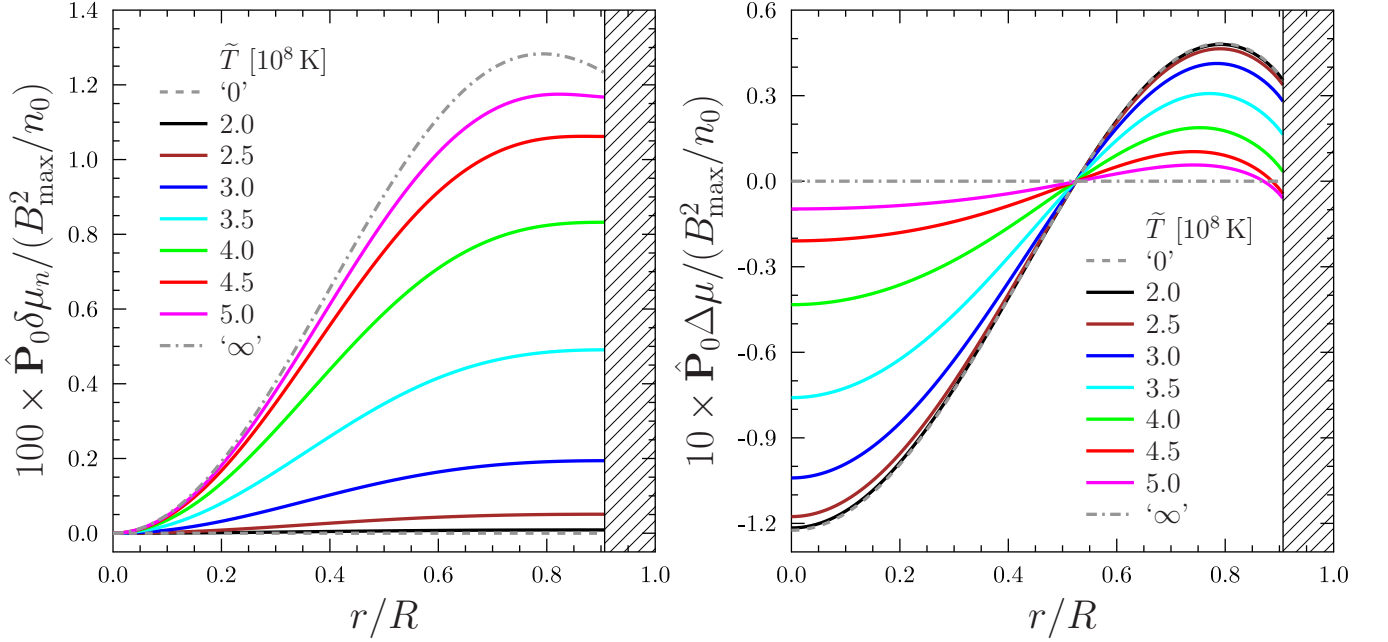}
	\caption{\label{fig:P0-min} 
		Radial dependence of the zeroth-order Legendre components 
		of (rescaled) chemical potential perturbations, $100 \times \delta\mu_n/(B_\text{max}^2/n_0)$ 
		and $10 \times \Delta\mu/(B_\text{max}^2/n_0)$, 
		for the minimal magnetic field model from Sec.\ \ref{sec:figures-minField}. 
		Colored lines show the exact solution to Eqs.~(\ref{eq:P0muFin}) 
		with the boundary conditions (\ref{eq:bound-cent}) and (\ref{eq:cc-SlowDiff}) 
		for a set of redshifted temperatures indicated in the figure. 
		Dashed lines 
		are plotted 
		in the low-temperature limit (almost coincide with the black lines);
		dot-dashed lines 
		illustrate the high-temperature limit. 
		Hatched regions correspond to the crust, where our approach is invalid.}
\end{figure}

Colored lines in Fig.~\ref{fig:P0-min} 
display solutions to Eqs.~(\ref{eq:P0muFin}) 
for the quantities $\PLeg_0\delta\mu_n$ and $\PLeg_0\Delta\mu$  
with the boundary conditions (\ref{eq:bound-cent}) and (\ref{eq:cc-SlowDiff}), 
for a number of redshifted stellar temperatures $\Tg$, listed in the figure.
For a sufficiently weak magnetic field, departure of the system 
from the beta-equilibrium state is small, 
$\Delta\mu \ll \pi k T$, hence 
one can present $\Delta \Gamma$ as \cite{GKO17}: 
$\Delta\Gamma = \lambda\Delta\mu$ (the subthermal regime).
Consider the system in this regime.
Then 
the 
differential equations (\ref{eq:P0muFin}) become linear, 
and, as a result, $\PLeg_0 \delta\mu_n$ and $\PLeg_0 \Delta\mu$ both proportional to $B_\text{max}^2$,
because $\PLeg_0 f_{Ar} \propto B_\text{max}^2$.
Thus, the dimensionless combinations 
$100 \times \PLeg_0 \delta\mu_n/(B_\text{max}^2/n_0)$ 
and $10 \times \PLeg_0 \Delta\mu/(B_\text{max}^2/n_0)$ 
($n_0 = 0.16\,$fm$^{-3}$ is the nuclear saturation density), 
which are shown in Fig.~\ref{fig:P0-min}, are independent of $B_\text{max}^2$.

Dashed and dot-dashed lines show, respectively,
the low-temperature and high-temperature limits 
for $\PLeg_0\delta\mu_n$ and $\PLeg_0\Delta\mu$ 
(see details in Sec.~\ref{sec:sol-bound}). 
One can see that the temperature $\Tg = 2\times 10^8\,$K 
is small enough to imitate the low-temperature limit 
(black lines in Fig.~\ref{fig:P0-min} almost coincide with the dashed ones). 
At $\Tg \sim 3.5\times 10^8\,$K 
the solution to Eqs.~(\ref{eq:P0muFin}) is equally far from both limits. 
The highest $\Tg$, for which we 
numerically solved Eqs.~(\ref{eq:P0muFin}), 
was $\Tg=5\times 10^8\,$K. 
This value is close to the high-temperature limit, 
but still significantly differs from it. 
At higher $\Tg$ Eq.\ (\ref{eq:P0muFin-long}) 
becomes poorly conditioned so that numerical integration of the system (\ref{eq:P0muFin}) 
is complicated.   

\begin{figure}
	\includegraphics[width=0.48\textwidth]{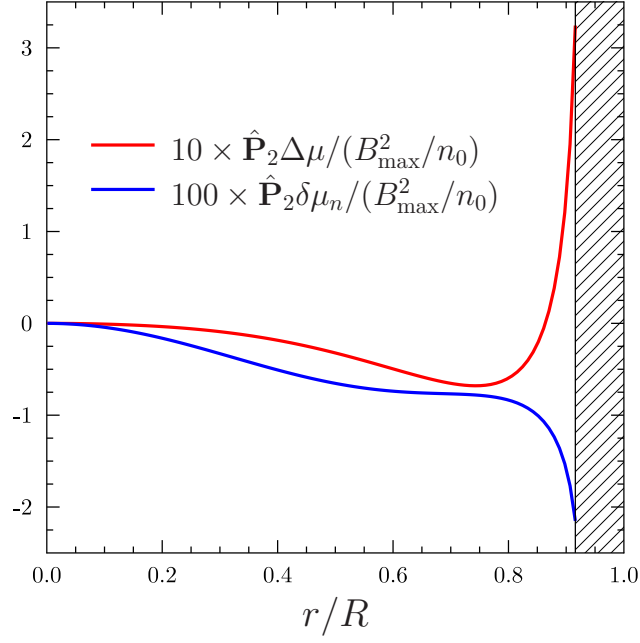}
	\caption{\label{fig:P2-min} 
		The same as in Fig.\ \ref{fig:P0-min}, but for the second-order Legendre components
		of $\delta\mu_n$ and $\Delta\mu$.}
\end{figure}

The flux function (\ref{eq:PsiThis}) contains two Legendre components, 
$P_0(\cos\theta)$ and $P_2(\cos\theta)$. 
The same is true for the chemical potential perturbations. 
Fig.~\ref{fig:P2-min} shows dimensionless combinations 
$100 \times \PLeg_2\delta\mu_n/(B_\text{max}^2/n_0)$ 
and $10 \times \PLeg_2\Delta\mu/(B_\text{max}^2/n_0)$ as functions of radial coordinate.
According to Eq.~(\ref{eq:DmudmuHighL}), 
these combinations do not depend on $B_\text{max}^2$ 
and are fully determined by the magnetic field.%
%
\footnote{
	\label{est}	
	As follows from Eq.~(\ref{eq:DmudmuHighL}), 
	one has the following estimate: 
	$\PLeg_l \delta \mu_n/\PLeg_l \Delta \mu \sim n_e/n_b \sim 0.1$, for $l \geqslant 1$.
	This estimate works well for $\PLeg_2$-components shown in Fig.\ \ref{fig:P2-min}.}
%
Typically, $\PLeg_2$-components are comparable to
$\PLeg_0$-components in the major part of the core.
However, there is a rapid (but finite) growth of 
$\PLeg_2\delta\mu_n$ and $\PLeg_2\Delta\mu$ near the crust-core interface.
The reason for that is a substantial softening of EOS at $n_b \lesssim n_0$, 
which results in large derivatives $n'_b$ and $n'_e$ 
in the vicinity of the crust-core boundary. 
This effect is generic (should be present for all EOSs), 
as is argued in Appendix~\ref{app:dndr}. 

Now let us discuss what happens if the system is in the regime, in which
$\Delta\mu \gtrsim \pi k T$ (suprathermal regime).
In this case $\Delta \Gamma$ is a polynomial in $\Delta\mu/\pi k T$ \cite{reisenegger95}.
As follows from Figs.~\ref{fig:P0-min} and \ref{fig:P2-min}, 
a typical absolute value of the chemical potential imbalance is 
$\sim (0.03-0.1)\times B_\text{max}^2/n_0$. 
Correspondingly, a typical magnetic field $B_\text{max}^\star$ 
for the transition of the system into the suprathermal 
regime is given by 
\begin{equation}
\label{eq:BmaxStar}
B_\text{max}^\star \sim (0.8-1.5)\times 10^{16} \sqrt{\frac{T}{10^8\,\text{K}}}~{\rm G}.
\end{equation}
Note that this equation depends on the local temperature $T$,
which is greater than the redshifted one. 
In the deepest layers of the core $T$ can be larger than $\Tg$ by a factor of $\sim 1.5$. 
Then for a typical magnetar temperature \cite{bl16}, 
$\Tg \sim 2\times 10^8\,$K, 
one can expect $B_\text{max}^\star > 10^{16}\,$G. 
Such a large field can be reached in magnetars, 
but in what follows we prefer 
to avoid nonlinearities in $\Delta \Gamma$ 
by choosing $B_\text{max} = 5\times 10^{15}\,$G.

\begin{figure}
	\includegraphics[height=0.331\textheight]{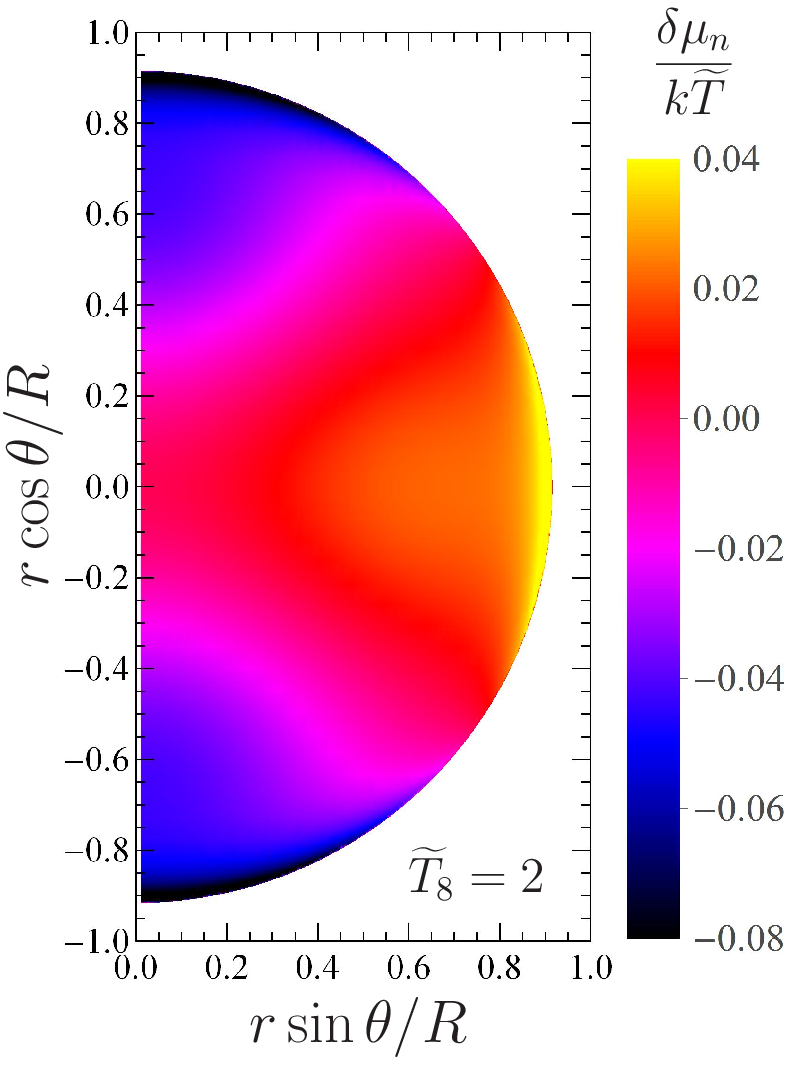}
	\includegraphics[height=0.331\textheight]{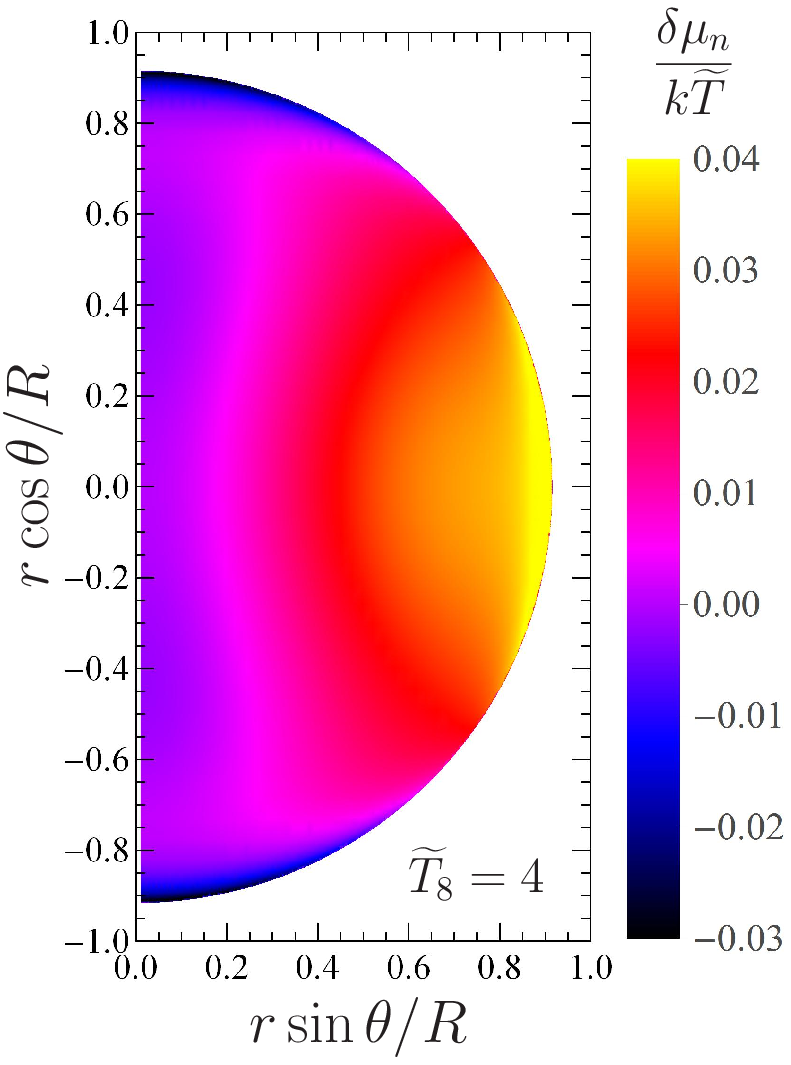}
	\includegraphics[height=0.331\textheight]{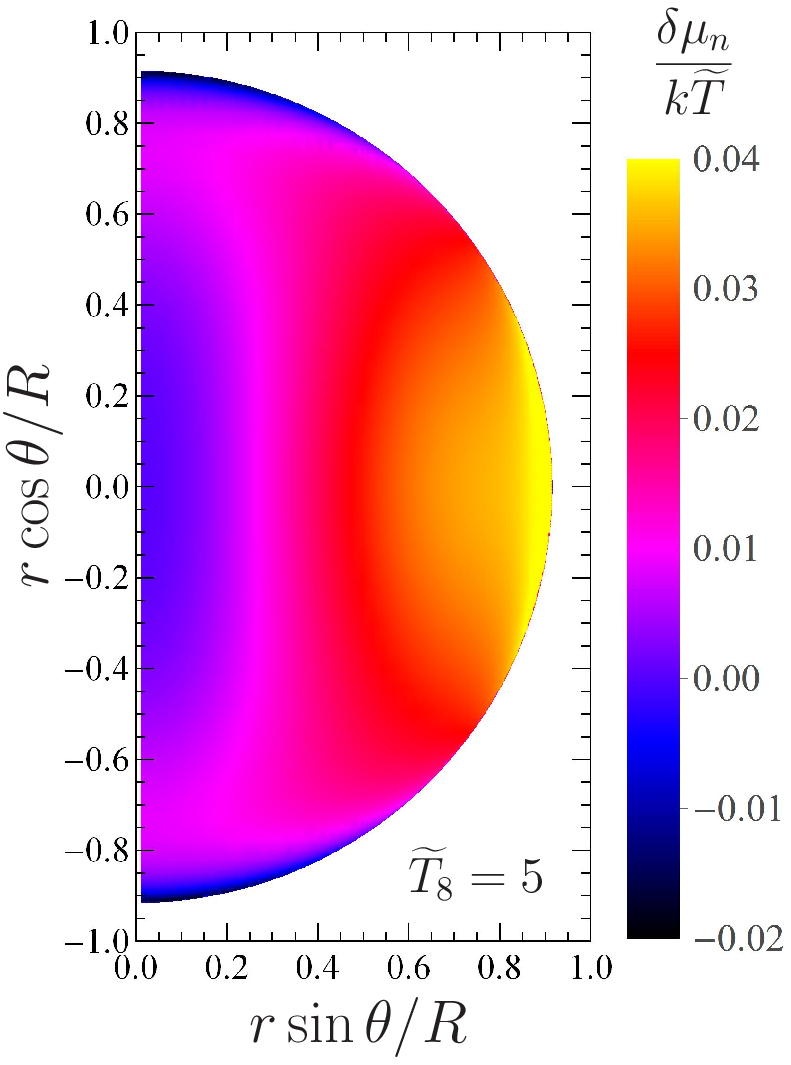}
	\caption{\label{fig:dmun-min} 
		Two-dimensional density plot showing the ratio of $\delta \mu_n/(k \Tg)$ in the star 
		for the minimal model of the magnetic field with
		$B_\text{max}=5\times 10^{15}\,$G. 
		The left, middle, and right panels 
		are plotted for $\Tg = 2\times 10^8\,$K, 
		$4\times 10^8\,$K,
		and $5\times 10^8\,$K, respectively.}
\end{figure}
\begin{figure}
	\includegraphics[height=0.334\textheight]{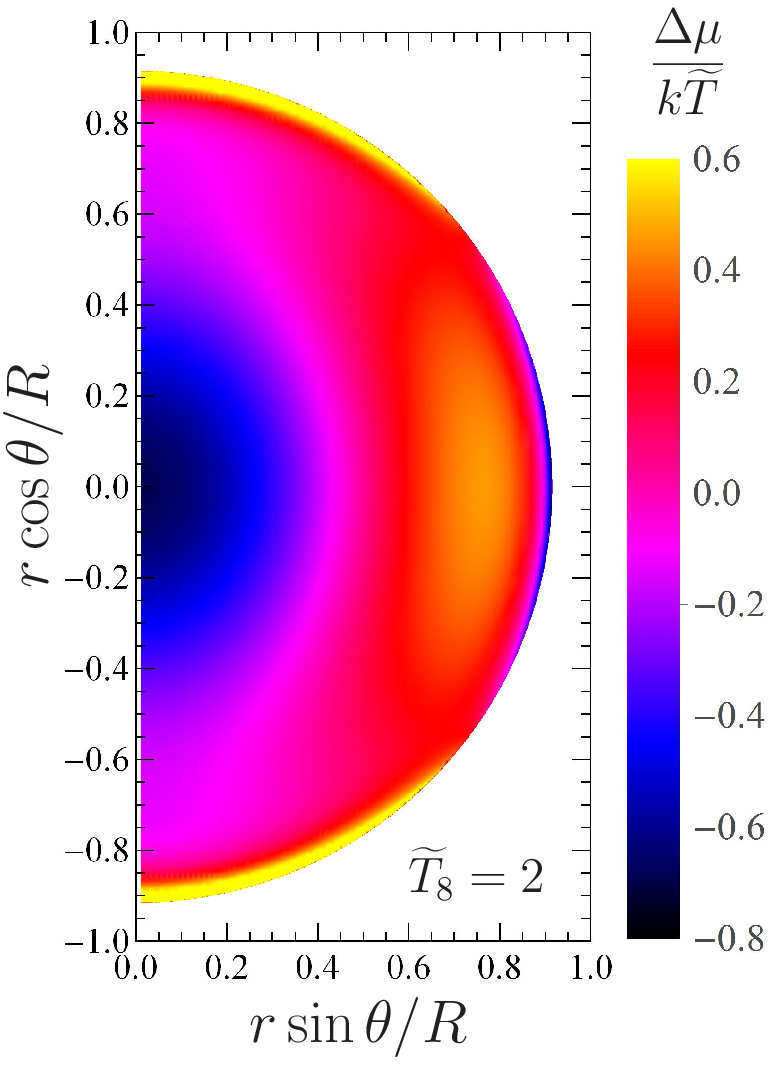}
	\includegraphics[height=0.334\textheight]{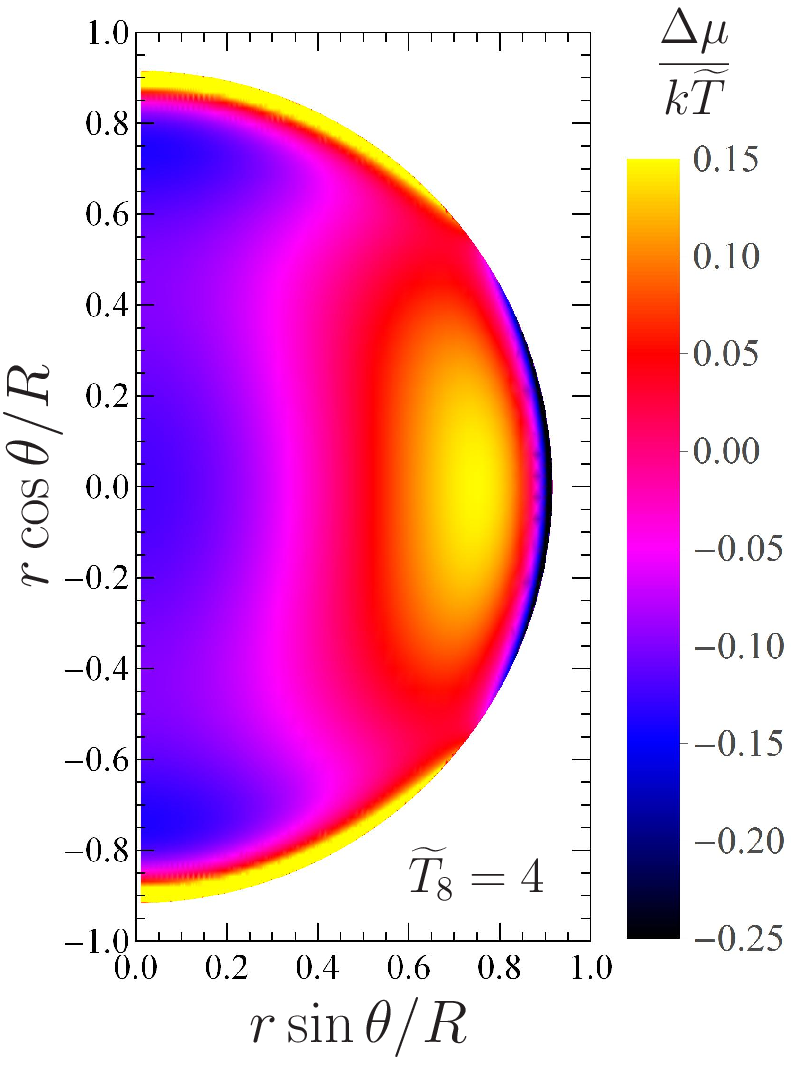}
	\includegraphics[height=0.334\textheight]{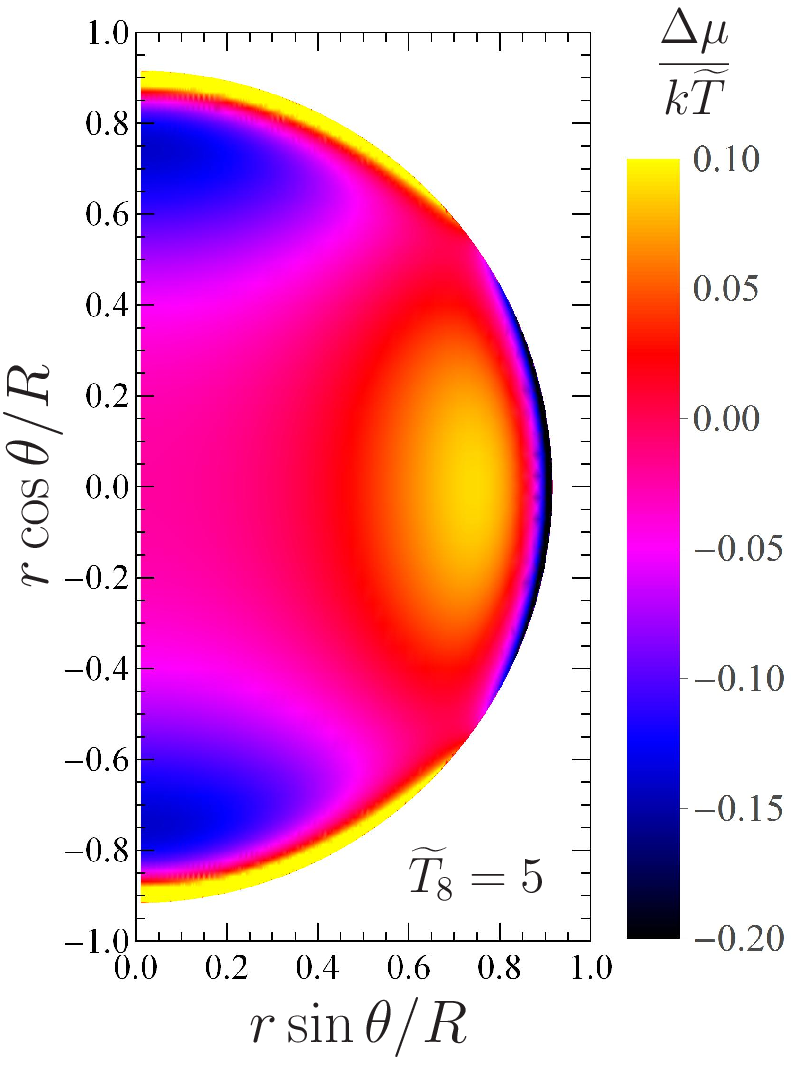}
	\caption{\label{fig:Dmu-min} 
		The same as in Fig.\ \ref{fig:dmun-min} but for the ratio $\Delta \mu/(k \Tg)$
		and $\Tg = 2\times 10^8\,$K (left panel), 
		$\Tg=4\times 10^8\,$K
		(middle panel), 
		and 
		$\Tg=5\times 10^8\,$K
		(right panel); see the text.}
\end{figure}

Our calculations are illustrated in Figs.~\ref{fig:dmun-min} and \ref{fig:Dmu-min}, 
where we present density plots for 
the ratios $\delta \mu_n/(k \Tg)$ and $\Delta \mu/(k\Tg)$. 
One sees from Fig.\ \ref{fig:dmun-min} that the second-order Legendre component 
of $\delta\mu_n$ dominates the zeroth one at 
$\Tg = 2\times 10^8\,$K; at $\Tg = 4\times 10^8\,$K 
and $5\times 10^8\,$K 
they are comparable, 
with a tendency that 
$\PLeg_0$-component becomes more and more important with increasing $\Tg$.
The situation with the chemical potential imbalance $\Delta\mu$ 
is reversed (Fig.\ \ref{fig:Dmu-min}). 
At $\Tg = 2\times 10^8\,$K the zeroth-order Legendre component, $\PLeg_0 \Delta \mu$, is slightlylarger than $\PLeg_2 \Delta \mu$,
but this difference smooths out at $\Tg = 4\times 10^8\,$K;
finally, at $\Tg=5\times 10^8\,$K the second-order component dominates.

\subsection{Poloidal flows in the minimal-field model}
\label{sec:figures-Uww}

\begin{figure}
\includegraphics[width=\textwidth]{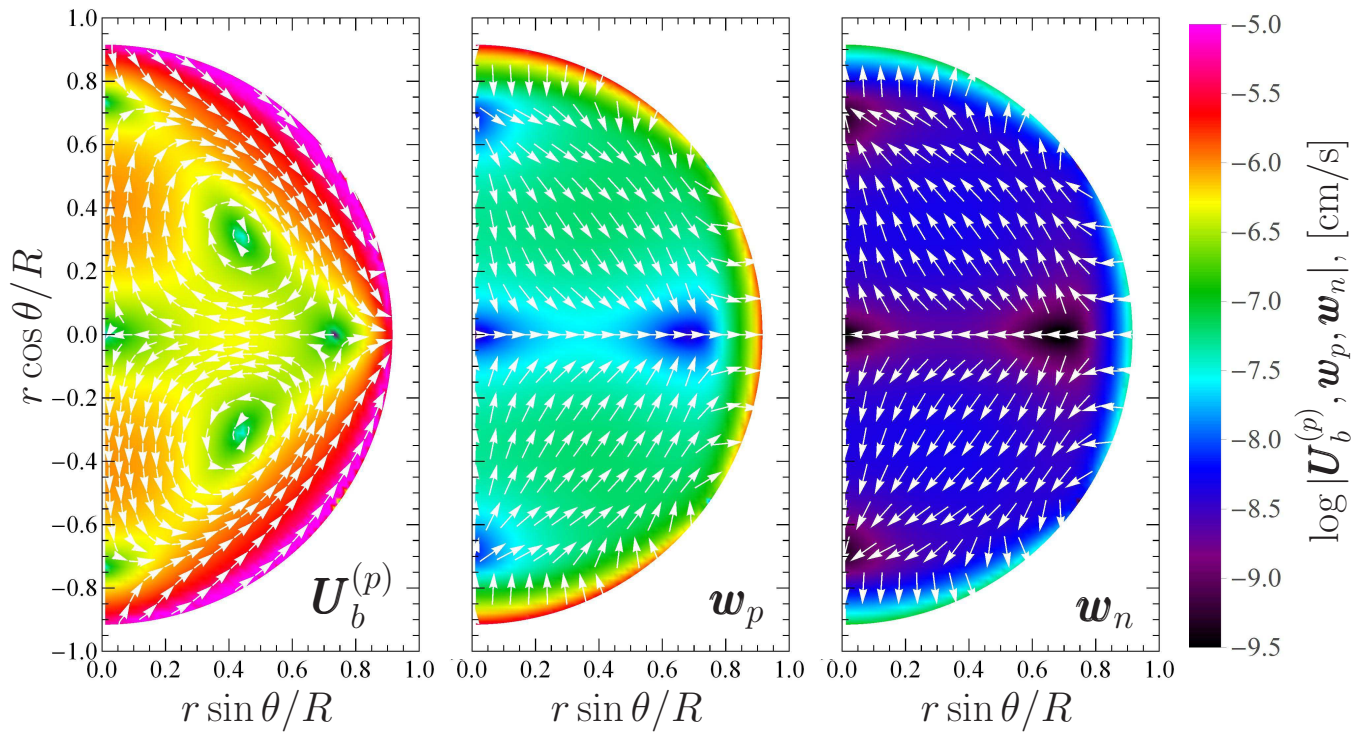}
\caption{\label{fig:UbWpWn-min} 
Two dimensional density plots showing 
the (logarithm of) poloidal component of the baryon velocity $\vec{U}_b^\pind$ (left panel),
the proton diffusion velocity $\vec{w}_p$ (middle panel), 
and the neutron diffusion velocity $\vec{w}_n$ (right panel)
for the minimal magnetic field model with $B_\text{max}=5\times 10^{15}\,$G 
and $\Tg=2\times 10^8\,$K. 
Arrows indicate the velocity direction.}
\end{figure}

In what follows we consider a neutron star with 
$\Tg = 2\times 10^8\,$K and $B_\text{max} = 5\times 10^{15}\,$G. 
The chosen $\Tg$ corresponds to the low-temperature limit, $\varkappa\gg1$, as 
discussed in Sec.\ \ref{sec:figures-mumu} (see the end of Sec.\ \ref{sec:sol-bound} for a definition of $\varkappa$). Then
one has $\PLeg_0 \mu_n=0$; in addition, the condition $\varkappa\gg1$ together 
with the estimate from the footnote \ref{est} allow us to neglect 
the rate $\Delta \Gamma$ in Eq.~(\ref{eq:Vdef}).
This significantly simplifies calculation 
of the poloidal component $\vec{U}_b^\pind$ of the baryon velocity, 
as well as the proton and neutron diffusion velocities,  
$\vec{w}_p$ and $\vec{w}_n$.

Fig.~\ref{fig:UbWpWn-min} presents these three quantities. 
Note that, for purely poloidal field $\vec{B}$ 
the poloidal projections of $\vec{w}_p$ and $\vec{w}_e$ coincide, 
while $w_{e\varphi} =- U_{b\varphi}$ (see Eq.\ \ref{wephi}).
One can see that the poloidal component $\vec{U}_b^\pind$, 
generated by the magnetic field, is rather high, 
$U_b^\pind \sim (0.3-1)\times 10^{-6}\,$cm/s,
and it rapidly increases near the crust-core interface.
It is also much larger than the typical diffusion velocities and the component $U_{b \varphi}$ (see Fig.~\ref{fig:Uphi-min}).%
%
\footnote{Note, however, that the property $U _b^\pind\gg U_{b \varphi}$ 
is a direct consequence of the poloidal nature 
of the adopted magnetic field model.
For a more realistic poloidal-toroidal configuration 
$U _b^\pind$ will be of the order of $U_{b \varphi}$
and, for example, the simple relation (\ref{wephi}) will no longer be satisfied.}
%
Typically, $\vec{U}_b^\pind$ in the core can be estimated as 
(for a given model of the magnetic field)
\begin{equation}
\label{eq:Ub-wp-wn}
U_b^\pind \sim (10-30) \, w_p, 
\end{equation}
while the proton and neutron diffusion velocities are related by the formula,
$\vec{w}_p=-(n_n/n_e) \vec{w}_n$ (see Eq.\ \ref{eq:Wsol}),
and thus $w_p\sim 10 w_n$.
Note that the relation (\ref{eq:Ub-wp-wn}) remains valid in a wide range of $\Tg$ 
(low enough to neglect $\Delta\Gamma$) and $B_\text{max}$, 
since all the three velocities  $\vec{U}_b^\pind$, $\vec{w}_p$, and $\vec{w}_n$
scale as $\propto B_\text{max}^2/\Tg$.%
%
\footnote{\label{Bnote}
It is interesting to note that in the superconducting matter 
these velocities will all be proportional to $B_{\rm max}$ rather than $B_{\rm max}^2$, 
because the magnetic force in that case is $\propto B_{\rm max}$ 
(see, e.g., Refs.\ \cite{gas11,gd16,GKO17}).}
%
The reason why $\vec{U}_b^\pind$ 
is much larger than $\vec{w}_p$ and $\vec{w}_n$ can be explained as follows. 
Let us express, for example, $\theta$-component, $U_{b \theta}$,
of the baryon velocity through $\vec{w}_p$.
Using Eqs.\ (\ref{eq:Wsol-p}), (\ref{eq:Vdef}), and (\ref{eq:Upfin}), 
one can write 
\begin{align}
U_{b \theta} =\frac{1}{n_b r \, {\rm sin}\theta} \, \frac{\partial}{\partial r} 
\left( \frac {r^2} {x_e'}\int_ {0}^\theta {\rm div} (n_e\vec {w}_p)\, {\rm sin}\tilde{\theta} \, d \tilde{\theta}
\right),
\label{Ubth}
\end{align}
where $x_e \equiv n_e/n_b$. Naive estimate of this expression, 
which assumes $d/dr \sim 1/R$,
suggests, incorrectly, that $U_{b \theta} \sim w_p$.
What is wrong here? Note that $\vec{w}_p$ depends on various derivatives of 
$n_b$ and $n_e$, as well as on the derivatives of
the magnetic function $f(r/R)$ (see Eqs.\ \ref{eq:Wsol-p} and \ref{eq:DmudmuHighL}),
namely, on $n_b'$, $n_e'$, $n_b''$, $n_e''$, $f$, $f''$, and $f'''$.                   
Correspondingly, $U_{b\theta}$ depends on up to the {\it fourth} derivative of $n_b$ and $n_e$ and up to {\it fifth} derivative of $f$ (see Eq.\ \ref{Ubth}).
These derivatives can be very large, so that, for example, 
an estimate $n_b''''(r) \sim n_b''(r)/R^2$ simply does not work; 
instead we have $n_b''''(r) \gg n_b''(r)/R^2$ in the outer layers of NS cores.
The reason for that is discussed in Appendix \ref{app:dndr},
where we, using as an example the $|n_b'/(n_b/R)|$ ratio,
demonstrate that the derivatives are especially large 
near the crust-core interface,               
where the stellar density rapidly decreases with $r$.
In turn, the fact, that $U_{b\theta}$ 
depends on up to the fifth derivative of 
the function $f$ indicates that baryon velocity can be extremely sensitive 
to the model of the magnetic field inside the NS core.
Note, in this respect, that the model (\ref{eq:PsiThis}), 
used by us here, is one of the smoothest 
(the magnetic field has a maximum at the stellar centre 
and then decreases monotonically in the direction of the crust, see Fig.~\ref{fig:B-min}).
We expect (and we checked it numerically for a number of different magnetic field configurations) 
that for less smooth models the baryon velocity will generally be even larger. 
Summarizing, the above consideration suggests that ${\pmb U}_b$ 
will exceed the diffusion velocities 
for a wide class of magnetic field geometries, 
except, perhaps, for some very specific configurations, 
in which large-derivative terms cancel each other out
(see also Sec.~\ref{disc}).

Since $\vec{U}_{b}$ dominates the diffusion velocities, 
all particle species move almost as a single fluid, 
$\vec{u}_a=\vec{U}_b+\vec{w}_a \approx \vec{U}_b$.
Using the expression (\ref{eq:Wsol-p}), 
it is easy to estimate various terms in Eq.\ (\ref{eq:EcomSol})  
and verify that the main contribution to $\delta\vec{E}_\text{com}$
comes from the third term in the right-hand side of Eq.\ (\ref{eq:EcomSol}),
so that $\delta E_\text{com} \sim (w_p/c) B \sim 0.05 |(\vec{U}_b^\pind/c) \times \vec{B}|$.
Correspondingly, the magnetic field lines 
are ``frozen-in'' to the $npe$-plasma to a good approximation 
(see Eq.\ \ref{eq:faradCom}),
\begin{equation}
\label{eq:faradCom1}
\pd{\vec{B}}{t} \approx
\rot\left( \vec{U}_b\times\vec{B} \right).
\end{equation}

An interesting feature of the solution presented in Fig.\ \ref{fig:UbWpWn-min}
is non-vanishing neutron and proton flows through the crust-core interface. 
Although these flows conserve total number of neutrons and protons
in the core and, generally, cannot be excluded on physical grounds,
it looks a bit strange that they appear inevitably in our scheme 
(for a given magnetic field configuration)
and we do not have enough freedom to ``kill'' them, e.g.,
by tuning in the boundary conditions at the interface.%
%
\footnote{Actually, we have just one {\it scalar} boundary condition (\ref{eq:cc-SlowDiff})
at the crust-core interface, which, however, is automatically satisfied 
in the low-temperature limit we are interested in here.
}
%
The question is what, effectively, plays 
the role of a ``boundary condition'' at the crust-core interface?
We believe it is the magnetic field itself, which, 
together with the quasistationarity assumption, 
encodes all information about the particle flows through the interface.
To illustrate this point, in the next section we build up a magnetic field model, which prevents baryons from moving into the crust.

\subsection{A magnetic field model with no baryon current through the crust-core interface}
\label{sec:figures-noFlow}

There is no baryon currents through the crust-core interface
if $u_{nr}|_{R_\text{core}} = u_{pr}|_{R_\text{core}} = 0$. 
This requirement can be reformulated in terms of the velocities $\vec{U}_{b}$, $\vec{w}_{p}$ and $\vec{w}_{n}$ as:
$U_{br}|_{R_\text{core}} = 0$ and $w_{pr}|_{R_\text{core}} = w_{nr}|_{R_\text{core}} = 0$. 
Employing Eqs.~(\ref{eq:Wsol-n}) and (\ref{eq:Wsol-p}), 
the latter condition can be rewritten as
\begin{equation}
\label{eq:noFlow-wCond}
\left( \PLeg_l \delta\mu_n \right)'\Bigr|_{R_\text{core}} =0, \quad l \geqslant 0.
\end{equation}
Note that Eq.\ (\ref{eq:noFlow-wCond})
for $l=0$  is automatically satisfied due to the
boundary condition (\ref{eq:cc-SlowDiff}). 
Using Eq.~(\ref{eq:Upfin}) for $\vec{U}_b^\pind$,
the condition $U_{br}|_{R_\text{core}} = 0$
is equivalent to $\PLeg_l V|_{R_\text{core}} = 0$ for each $l\geqslant 0$ 
(for $l=0$ this condition should already be satisfied, 
as explained in Sec.~\ref{sec:sol-mu}). 
In the low-temperature limit we are interested in here, 
we may omit the reaction rate $\Delta\Gamma$ in Eq.~(\ref{eq:Vdef}) for $V$. 
Then the condition $\PLeg_l V|_{R_\text{core}} = 0$ can be represented as 
\begin{equation}
\label{eq:noFlow-UCond}
\left[ \left( \PLeg_l\delta\mu_n \right)'' - \frac{l(l+1)}{r^2} \PLeg_l\delta\mu_n \right]\Biggr|_{R_\text{core}} = 0, \quad l \geqslant 1,
\end{equation}
where we have used Eq.~(\ref{eq:noFlow-wCond}).
Substituting the expressions 
(\ref{eq:DmudmuHighL}) for $\delta\mu_n$ into Eqs.~(\ref{eq:noFlow-wCond}) and (\ref{eq:noFlow-UCond}), we arrive at the two conditions for the  
Ampere force, $\vec{f}_A$. Using Eq.~(\ref{eq:app:fA-Psi}), 
these conditions can be written in terms of the flux and current 
functions, $\Psi$ and $I$.

\begin{figure}
\includegraphics[height=0.343\textheight]{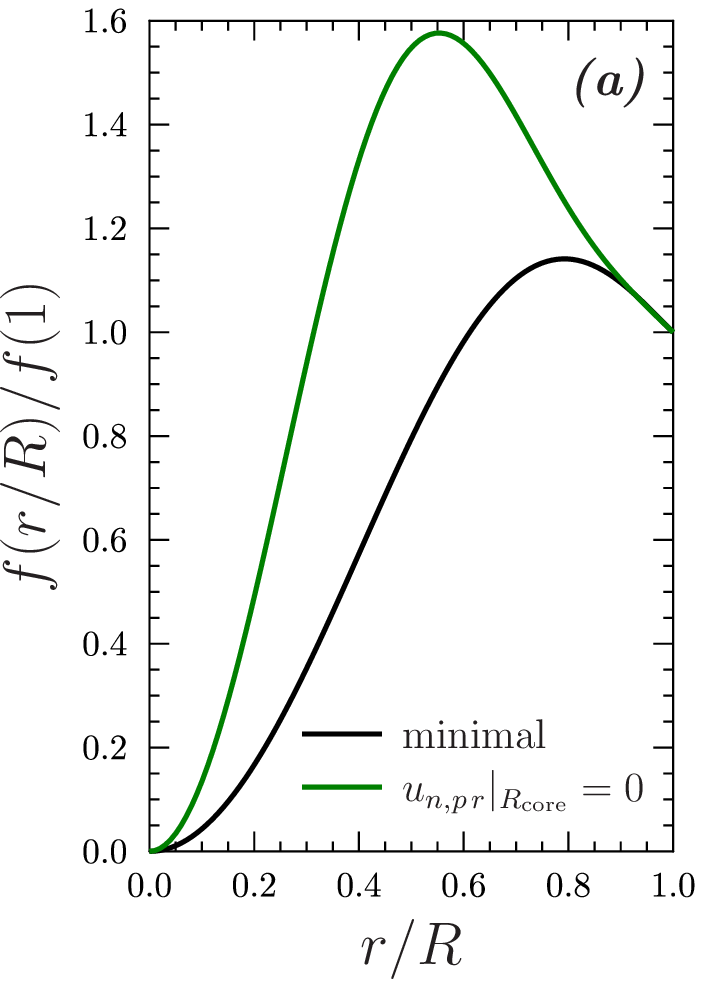}
\includegraphics[height=0.343\textheight]{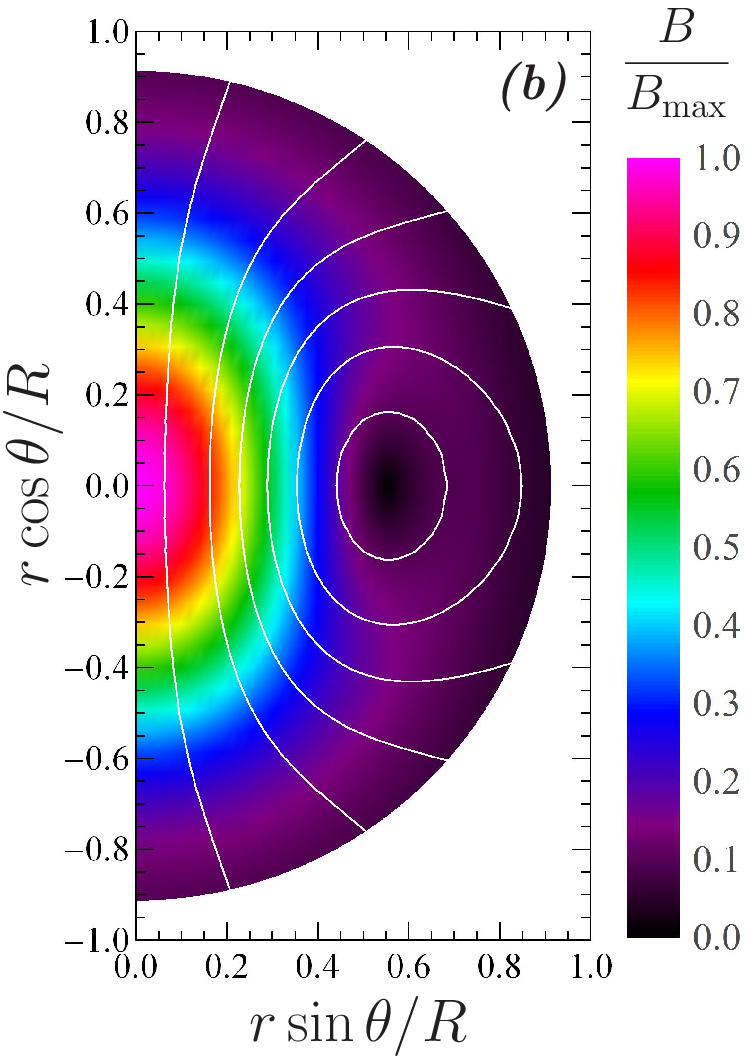}
\includegraphics[height=0.343\textheight]{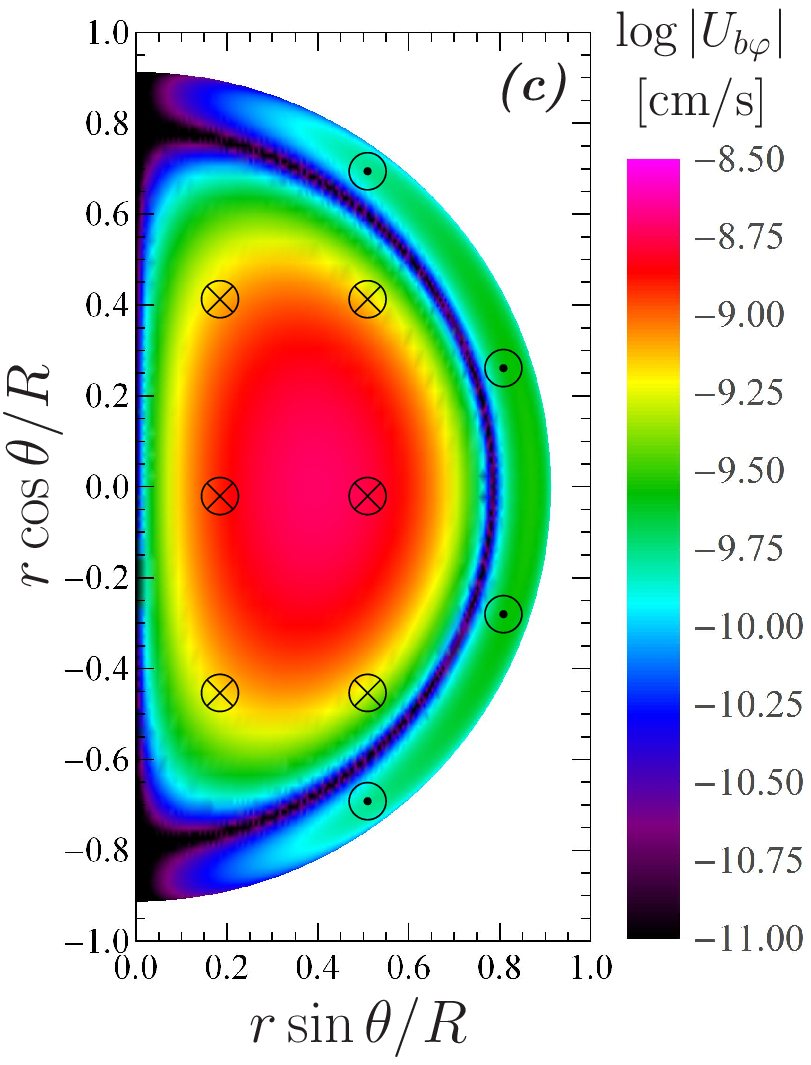}
\caption{\label{fig:fBUphi-noFlow} 
\textit{(a)} Comparison of 
the dimensionless functions $f$ 
for the minimal magnetic field model from Sec.~\ref{sec:figures-minField} (black line)
and for the model of Sec.~\ref{sec:figures-noFlow} (green line).
\textit{(b)} Magnetic field model from Sec.~\ref{sec:figures-noFlow} 
(cf. Fig.\ \ref{fig:B-min}). 
\textit{(c)} Density plot for the toroidal component of the baryon velocity 
for the magnetic field model 
from Sec.~\ref{sec:figures-noFlow} (cf. Fig.~\ref{fig:Uphi-min}). 
Circles with crosses and dots indicate direction of the toroidal component.}
\end{figure}
\begin{figure}
\includegraphics[width=\textwidth]{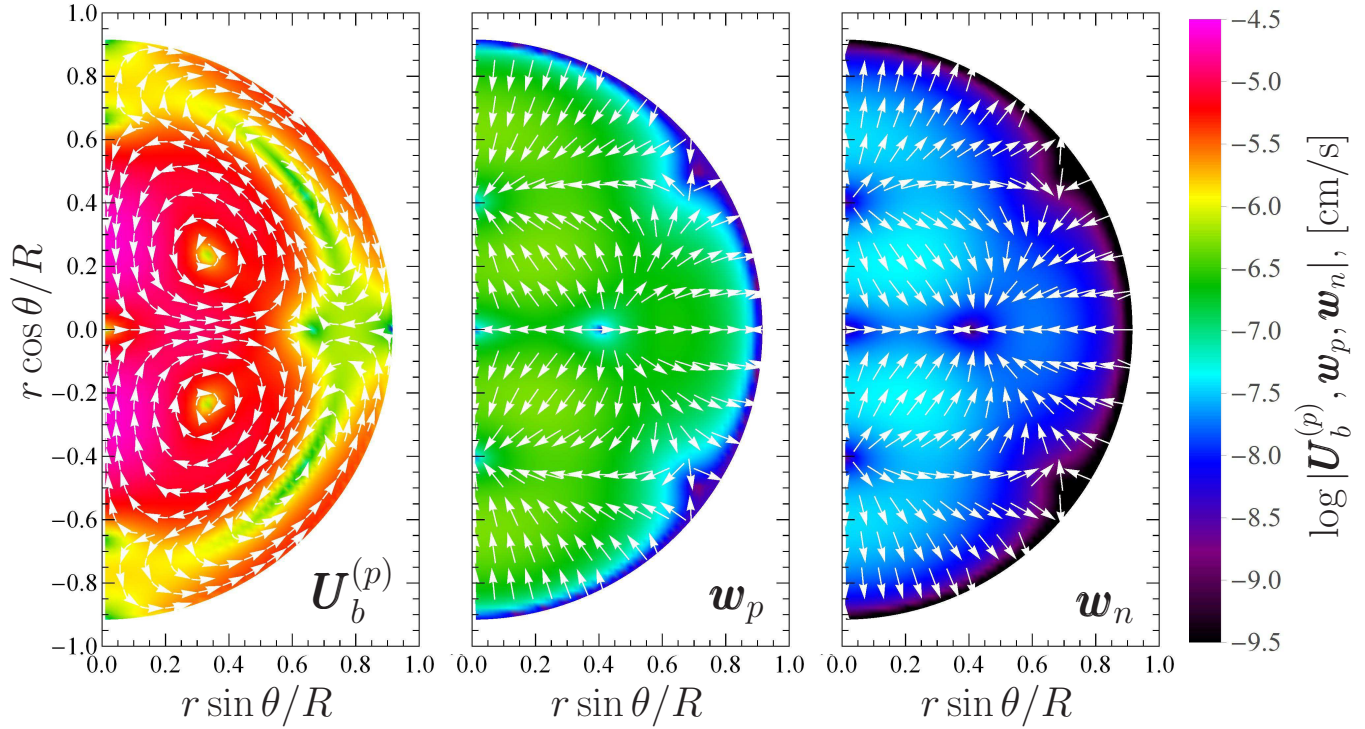}
\caption{\label{fig:UbWpWn-noFlow} 
The same as in Fig.~\ref{fig:UbWpWn-min}, but for the magnetic field model 
from Sec.~\ref{sec:figures-noFlow}}
\end{figure}

In the case of a purely poloidal field 
with simple angular dependence (\ref{eq:PsiThis}) of the function $\Psi$, 
the conditions (\ref{eq:noFlow-wCond}) and (\ref{eq:noFlow-UCond}) are nontrivial only for the Legendre component with $l=2$. 
Thus, we have two new conditions for the function $f(x)$ and the three old conditions, namely Eq.~(\ref{eq:fBounds-surf}) and the normalization. 
A simple (numerical) solution, satisfying these conditions, reads 
(five coefficients in the formula are needed
to satisfy 5 conditions)
\begin{equation}
\label{eq:f-noFlow}
f = 0.5 x^2 - 1.557 x^4 + 2.113 x^6 - 1.358 x^8 + 0.338 x^{10}.
\end{equation}
In Fig.~\ref{fig:fBUphi-noFlow}a we compare this function 
with $f(x)$ for the minimal-field model (see Eq.~\ref{eq:f-min}).
Both models look similar, as is also confirmed by
Fig.~\ref{fig:fBUphi-noFlow}b, presenting the new magnetic field model.
It resembles Fig.~\ref{fig:B-min}
with the only difference that now $\vec{B}$ 
is more ``compressed'' to the central regions of the core.
The main difference concerns the 
toroidal component of the baryon velocity (Fig.~\ref{fig:fBUphi-noFlow}c),
which has a more complicated topology comparing to $U_{b\varphi}$ 
for the minimal model of the magnetic field
(Fig.~\ref{fig:Uphi-min}). 

Fig.~\ref{fig:UbWpWn-noFlow} shows the baryon 
and diffusion velocities for the new model of the magnetic field. 
We see that this field indeed suppresses all radial velocities 
at the crust-core interface. 
However, the $U_{b\theta}$ component near $r=R_\text{core}$ does not vanish. 
The structure of the flows significantly differs
from what we see in Fig.~\ref{fig:UbWpWn-min}. 
In particular, we have four curls in Fig.~\ref{fig:UbWpWn-noFlow} (left panel)
instead of two, as in the minimal-field model.
Average absolute values of the velocities 
are plotted in Fig.~\ref{fig:UbWpWn-noFlow} 
and appear to be about one order of magnitude larger 
than in Fig.~\ref{fig:UbWpWn-min}.
However, the scaling relation (\ref{eq:Ub-wp-wn}) remains unaltered.

\section{Discussion \& Conclusions}
\label{disc}

In our previous paper \cite{GKO17} 
we proposed a self-consistent method to explore 
the magnetic field evolution in NS cores. 
The main attractive feature of the method is that
it does not assume, from the very beginning, 
that the neutron or baryon velocities are small or vanish 
(as, e.g., in Refs.~\cite{gr92,td96,gas11,gagl15,papm17}),
but instead provide a receipt to calculate them.
In the present paper we further developed this method and applied it
to the model of a nonsuperfluid and nonsuperconducting NS, 
whose core consists of neutrons, protons, and electrons.
Such a NS model may adequately describe
hot, high-$B$ magnetars with
partially or fully suppressed baryon superfluidity in their cores \cite{ss15,sxsc17}.
Our main result is the calculation of particle velocities 
in the NS core, generated by the presence of the magnetic field. 

\begin{figure}
	\includegraphics[width=0.45\textwidth]{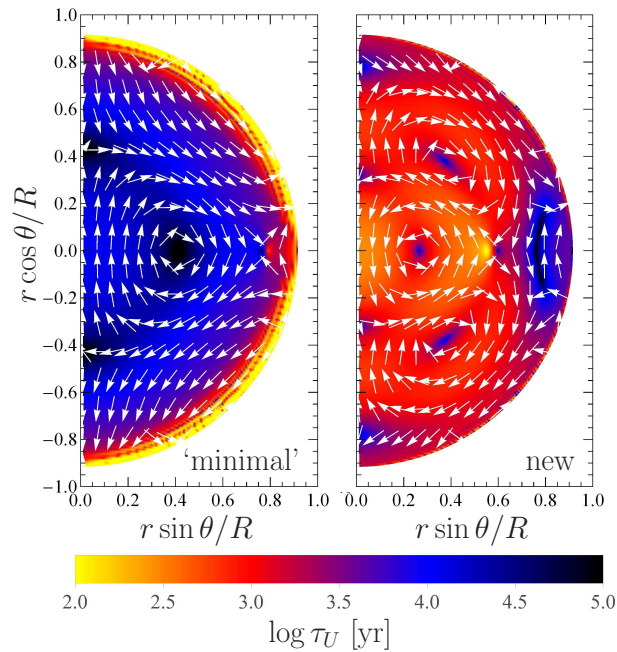}
	\caption{\label{timescale}
		Density plot showing the logarithm of the 
		timescale $\tau_U$ 
		(in yr) 
		for the minimal model of the magnetic field from Sec.~\ref{sec:figures-minField} (left panel) and for the model
		with no particle flows through the crust-core interface 
		from Sec.~\ref{sec:figures-noFlow} (right panel). 
		Arrows indicate direction of the derivative $\partial \vec{B}/\partial t$ given by Eq.~(\ref{eq:faradCom1}).
		To plot the figure we assumed $B_{\rm max}=5\times 10^{15}$~G and $\Tg=2 \times 10^8$~K.
	}
\end{figure}
%

There are several interesting properties of our solution
which, as we argue in Sec.~\ref{sec:figures-Uww}, should be common to 
a wide class of magnetic-field models (see also a discussion below).
First and most importantly, we showed that neutrons, protons, and electrons 
in the core move almost as a single fluid, i.e., the
diffusion velocities $\vec{w}_a$ ($a=n$, $p$, $e$) 
are much smaller than the baryon velocity $\vec{U}_b$ 
(see Eq.~\ref{eq:UWdef}). This result contradicts the statement 
that neutrons (or baryons) in NSs are motionless to 
a good approximation (e.g., Refs.~\cite{gr92,td96,bl16,papm17,crv2017}).
Specifically, we showed, for the two models of the magnetic field 
discussed in Sec.~\ref{sec:figures},
that for charged particles 
$w_{p,e} =|\vec{u}_{p,e}-\vec{U}_b|\sim 0.05 U_b$, 
while for neutrons $w_n=|\vec{u}_n-\vec{U}_b|\sim 0.005 U_b$.
Note that here the factors in front of $U_b$ 
depend sensitively on the magnetic field model and
can be made substantially smaller for a less ``smooth'' magnetic function 
$f(x)$ (see Sec.~\ref{sec:figures-Uww}).

The second property is that, typically, the velocity $\vec{U}_b$ is rather high
and it rapidly increases in the outer layers of the core.
For example, $U_b \sim (0.3-1)\times 10^{-6}$~cm/s 
for the magnetic field model with $B_\text{max} = 5\times 10^{15}$~G and
temperature $\Tg = 2\times 10^8$~K. On the other hand, the particle velocities 
$\vec{u}_a$ and $\vec{U}_b$ scale similarly with $B_\text{max}$ and 
$\Tg$ in the low- and high-temperature regimes%
%
\footnote{For example, in the low-temperature regime, $\Tg \lesssim 3\times 10^8$~K, $u_a \propto B_\text{max}^2/\Tg^2$.
}
%
(see Sec.~\ref{sec:figures-mumu}), thus $\vec{U}_b$ will 
remain much larger than the diffusion velocities $\vec{w}_a$
if this is the case for some particular values of $B_{\rm max}$ and $\Tg$.%
%
\footnote{
Note that the second term in the expression (\ref{eq:Wsol-e}) for 
the electron diffusion velocity, $\vec{w}_e$,  
depends on the charge current density $\vec{j}$ and hence $\propto B_{\rm max}$ 
rather than $\propto B^2_{\rm max}$.
However, this term is very small for magnetars (see Sec.\ \ref{sec:figures-Uww}).
At the same time, for ordinary pulsars all the diffusion velocities, 
as well as the baryon velocity $\vec{U}_b$ scale as $\propto B_{\rm max}$ 
(see footnote \ref{Bnote}), 
so that $\vec{w}_e$ will be much smaller than $\vec{U}_b$ at arbitrary $\vec{B}$.
}
%

Another interesting property of our solution 
is that some magnetic field configurations can lead 
to neutron and proton flows through the crust-core interface. 
Although, generally, we see no reasons why such behavior is impossible, 
we derived a set of conditions (Sec.~\ref{sec:figures-noFlow}) 
on the magnetic field, preventing these particle species from 
penetrating the crust, and found
a magnetic field model, satisfying these requirements. 
Note that the ``minimal'' model of Refs.\ \cite{papm17,armm13} 
is not compatible with these requirements.

The solution that we have just discussed was obtained for two specific models of the magnetic field. But how representative are these models?
In Sec.~\ref{sec:figures-Uww}
we argue that the main feature of our solution ---
large baryon velocity ${\pmb U}_b$, exceeding the diffusion velocities ${\pmb w}_a$ --- is inherent in a wide class of magnetic-field models 
[see a discussion after Eq.~(\ref{Ubth})].
This is because ${\pmb U}_b$ depends on high spatial derivatives 
of the magnetic field and number densities, which can be very large.
Of course, one can imagine a ``fine-tuned'' model 
(or a set of models), in which these large derivatives 
cancel each other out, so that the resulting ${\pmb U}_b$ 
is comparable to the diffusion velocities.
But what is the probability that such a model is realized in an NS, for instance, at its birth? The magnetic field of a newly-born 
star is determined by the way it is generated in the process of NS formation.
It knows nothing about what model leads or does not lead to large ${\pmb U}_b$.
Therefore, we expect that the probability to have a ``fine-tuned'' 
magnetic-field configuration for a young NS is negligible.

However, one cannot exclude that the magnetic field may converge
to the ``fine-tuned'' configuration during the NS evolution.
This will happen if the models with $U_b\sim w_a$ constitute an attractor for our problem, in analogy with the Hall attractor in the NS crust \cite{gc14}.
Note that, for this scenario to work, 
the magnetic field configuration corresponding to the attractor state
should be dynamically stable. Otherwise, approaching the attractor, 
the magnetic field will reconfigure on the Alfven timescale
to some other (stable) configuration 
(most likely, with large ${\pmb U}_b$), 
and the cycle of quasistationary evolution starts over again.
Whether these scenarios are realized in reality
is an open question that needs further investigation. 

If the configuration of the magnetic field is such that
the NS matter moves almost as a single fluid (i.e., $U_b>w_a$),
then ${\pmb B}$ is (approximately) frozen-in to the $npe$-plasma,
so that the field lines are dragged with it
[see Eq.\ (\ref{eq:faradCom1})].
This introduces a new timescale into the problem
(not to be confused with the dissipation timescale studied in 
Ref.\ \cite{GKO17}!), associated with the drag velocity ${\pmb U}_b$,
$\tau_U = B/|\rot (\vec{U}_b\times \vec{B}) |$.
The timescale is plotted in Fig.~\ref{timescale} for the two 
models of the magnetic field discussed in the text.
To plot the figure we assumed $B_{\rm max}=5\times 10^{15}$~G and 
$\Tg=2 \times 10^8$~K, but the result can be easily rescaled to any 
$B_{\rm max}$ and $\Tg$ since 
$\tau_U \propto B_{\rm max}^2/\Tg^2$ for $\Tg \lesssim 3 \times 10^8$~K.
One sees, that for the minimal magnetic field model (left panel) 
$\tau_U \sim 10^4$~yr in the bulk of the core,
but becomes much smaller, $\lesssim 10^2$~yr, in the vicinity of the crust. 
The behavior of $\tau_U$ for the model with suppressed particle flows 
through the crust-core boundary (right panel) 
is quite different: $\tau_U$ for that model is $\sim 10^3$~yr in the major part of the core. It seems plausible that the real timescale $\tau$ 
for the magnetic field evolution in the star 
will be closer to the minimal value of $\tau_U$, i.e., 
$\tau \sim 10^2$~yrs for the model of Sec.~\ref{sec:figures-minField} 
and $\tau\sim 10^3$~yrs for the model of Sec.~\ref{sec:figures-noFlow}.
In both cases $\tau$ is noticeably smaller than the typical magnetar age, 
$\sim 10^4$~yr \cite{ok14}.%
%
\footnote{
This timescale can be further reduced by choosing 
a less smooth model of the magnetic field. 
}
%
This observation suggests 
a principal possibility that
reconfiguration of the magnetic field 
in the core may be 
responsible for violent magnetar activity during its lifetime,
or even may lead to (repeating) fast radio bursts \cite{pp13,beloborodov17,ppp18}.%
%
\footnote{The idea that rapid evolution of the magnetic field in the core
may explain repeating FRBs has been communicated to us by S.B.~Popov.}
%

Of course, we are still far from the final solution 
to the problem of the magnetic field evolution in NS interiors. 
Now we are able just to find all the necessary ingredients 
in order to calculate the derivative $\partial \vec{B}/\partial t$ 
for a simplest model of an NS  and for a predefined model of the magnetic field.
The next steps, therefore, could be 
(i) to solve the Faraday's equation to see how the magnetic field 
evolves in time and (ii) to extend the results obtained here to 
more realistic (and complex) models of NSs.
To reach the goal (i) one needs, first of all, 
to include the crust into consideration, 
to obtain a quasistationary solution for the magnetic field there, 
and then to match both solutions in the crust and core,
allowing for possible particle currents through the crust-core interface.
In turn, the goal (ii) assumes a lot of work.
One needs to take into account
general relativity effects in MHD equations
(which should not affect the results quantitatively),
to consider various core compositions (not only $npe$-matter), 
but the most important task will be to extend our 
consideration to superfluid and superconducting matter.
Such an extension is absolutely necessary for modeling 
radio and millisecond pulsars, and it can be done 
along the lines reviewed in Ref.\ \cite{GKO17}.
Our preliminary results indicate that account for nucleon pairing 
has a tendency to additionally reduce
the typical magnetic timescale $\tau_U$
in comparison to nonsuperfluid NSs. A detailed 
analysis of this problem  will be given in our forthcoming publication.

\begin{acknowledgments}
%
We are grateful to A.M.~Beloborodov, A.I.~Chugunov, 
E.M.~Kantor, S.B.~Popov, and D.G.~Yakovlev
for discussions and interest in our work.
This work is supported in part by the Foundation 
for the Advancement of Theoretical Physics and Mathematics BASIS 
[Grants No. 17-12-204-1 (MEG) and 17-15-509-1 (DDO)].
\end{acknowledgments}

\appendix

\section{Derivation of the evolution equation for the poloidal flux function}
\label{app:dtPsi}

As explained in Sec.~\ref{sec:sol-Uphi}, 
the axisymmetric magnetic field can be decomposed in terms of 
the poloidal flux function $\Psi$ and the current function $I$,%
%
\footnote{One should bear in mind that the gradients of these functions
and, generally, 
the gradients of any scalars in our problem
are poloidal vectors, 
which are perpendicular to $\vec{\nabla}\varphi$. 
This fact is actively used in Appendices \ref{app:dtPsi}, \ref{app:Uphi}, 
and in Sec.~\ref{sec:sol-Uphi}.}
%
%
\begin{equation}
\label{eq:app:Bpt}
\vec{B} = \vnabla\Psi\times\vnabla\varphi + I\vnabla\varphi = \rot\left( \Psi\vnabla\varphi \right) + I\vnabla\varphi,
\end{equation}
where $\vnabla\varphi = \vec{e}_\varphi/(r\sin\theta)$. 
Then the electric current density is
\begin{equation}
\label{eq:app:jPsi}
\vec{j} = \frac{c}{4\pi}\rot\vec{B} = -\frac{c}{4\pi}\grsh\Psi\vnabla\varphi + \frac{c}{4\pi}\vnabla I\times\vnabla\varphi,
\end{equation}
where
\begin{equation}
\label{eq:app:GSdef}
\grsh = \Delta + \frac{\vec{\nabla}\left[ (\vec{\nabla}\varphi)^2 \right]}{(\vec{\nabla}\varphi)^2}\cdot\vec{\nabla} = \pd{^2}{r^2} + \frac{\sin\theta}{r^2}\pd{}{\theta}\left( \frac{1}{\sin\theta}\pd{}{\theta} \right)
\end{equation}
is the Grad-Shafranov operator (see, e.g., Ref.~\cite{Goed2010}). 
The Ampere force $\vec{f}_A$ from Eq.~(\ref{eq:MHDquas-totEuler}) takes the form 
\begin{equation}
\label{eq:app:fA-Psi}
\vec{f}_A = -\frac{(\vec{\nabla}\varphi)^2}{4\pi}\left( \grsh\Psi\vec{\nabla}\Psi + I\vec{\nabla}I + \vec{\nabla}\Psi\times\vec{\nabla}I \right).
\end{equation}
Due to an axisymmetry there is no forces to balance $f_{A\varphi}$, 
i.e., Eq.~(\ref{eq:fAtor}) is satisfied and, consequently, $I = I(\Psi,t)$. 
Eq.~(\ref{eq:faradSplit-pol}) can be rewritten as
\begin{equation}
\label{eq:app:dtPsi-rotrot}
\rot\left( \pd{\Psi}{t}\vec{\nabla}\varphi \right) = \rot\left[ -c\delta\vec{E}_\text{com}^\tind + \vec{U}_b^\pind\times\left( \vec{\nabla}\Psi\times\vec{\nabla}\varphi \right) \right],
\end{equation}
therefore,
\begin{equation}
\label{eq:app:dtPsi-tmp}
\pd{\Psi}{t}\vec{\nabla}\varphi = -c\delta\vec{E}_\text{com}^\tind - \vnabla\varphi\left( \vec{U}_b^\pind\cdot\vnabla\Psi \right) + \vec{\nabla}\eta,
\end{equation}
were $\eta$ is an arbitrary scalar function. 
It does not affect the evolution of observable quantities and can be omitted.
Moreover, since the system is axisymmetric, 
we can treat $\eta$ as a function of only $r$ and $\theta$, 
thus it has to vanish since the other terms in Eq.~(\ref{eq:app:dtPsi-tmp}) 
are purely toroidal.

Employing Eqs.~(\ref{eq:EcomSol}) and (\ref{eq:app:Bpt}), we have
\begin{equation}
\label{eq:app:Ecomt}
\delta\vec{E}_\text{com}^\tind = \frac{J_{ep}}{e^2 n_e^2}\vec{j}^\tind - \frac{n_n^2}{c n_b J_{np}}\vnabla\delta\mu_n\times\vec{B}^\pind = -\frac{c J_{ep}}{4\pi e^2 n_e^2}\grsh\Psi\vnabla\varphi + \frac{n_n^2}{c n_b J_{np}}\left( \vnabla\delta\mu_n\cdot\vec{\nabla}\Psi \right)\vnabla\varphi.
\end{equation}
Using this equation together with Eq.~(\ref{eq:app:dtPsi-tmp}), 
we finally obtain,
\begin{equation}
\label{eq:app:dtPsi-fin}
\pd{\Psi}{t} = \frac{c^2 J_{ep}}{4\pi e^2 n_e^2}\grsh\Psi - \frac{n_n^2}{n_b J_{np}}\vnabla\delta\mu_n \cdot\vec{\nabla}\Psi - \vec{U}_b^\pind\cdot\vnabla\Psi.
\end{equation}
This equation 
coincides with 
Eq.~(\ref{eq:PsiEvol}).

\section{Derivation of Eq.~(\ref{eq:It})}
\label{app:Uphi}

Multiplying Eq.~(\ref{eq:dtIDphi}) by $\vnabla\varphi$, one has
\begin{multline}
\label{eq:app:dtI=div}
\left( I'_\Psi\pd{\Psi}{t} + I'_t \right)(\vnabla\varphi)^2 = \vnabla\varphi\cdot \rot\left( -c\delta\vec{E}_\text{com}^\pind + I\vec{U}_b^\pind\times\vnabla\varphi + \vec{U}_b^\tind\times\vec{B}^\pind \right) =\\
= \diver\left[ - c\delta\vec{E}_\text{com}^\pind\times\vnabla\varphi - I (\vnabla\varphi)^2\vec{U}_b^\pind + \vec{B}^\pind\left( \vnabla\varphi\cdot\vec{U}_b^\tind \right) \right].
\end{multline}
The last term in the right-hand side of this equation can be rewritten as
\begin{equation}
\label{eq:app:div3}
\diver\left[ \vec{B}^\pind\left( \vnabla\varphi\cdot\vec{U}_b^\tind \right) \right] = \vec{B}^\pind\cdot\vnabla\left( \vnabla\varphi\cdot\vec{U}_b^\tind \right),
\end{equation}
since $\diver\vec{B}^\pind = 0$. 
The middle term in the right-hand side of Eq.~(\ref{eq:app:dtI=div}) equals
\begin{equation}
\label{eq:app:div2}
-\diver\left[ I(\vnabla\varphi)^2\vec{U}_b^\pind \right] = -I\diver\left[ (\vnabla\varphi)^2\vec{U}_b^\pind \right] - (\vnabla\varphi)^2 I'_\Psi \vec{U}_b^\pind\cdot\vnabla\Psi,
\end{equation}
where we make use of Eq.~(\ref{eq:Ideriv-r}).
The poloidal component of the comoving electric field (\ref{eq:EcomSol}) is
\begin{equation}
\label{eq:app:Ecomp}
\delta\vec{E}_\text{com}^\pind = \frac{J_{ep}}{e^2 n_e^2}\vec{j}^\pind - \frac{n_n^2}{c n_b J_{np}}\vnabla\delta\mu_n\times\vec{B}^\tind + \frac{\vec{f}_A^\pind}{e n_e}.
\end{equation}
Here we omit the term $-\vnabla\delta\mu_e/e$ 
since it does not contribute to $\rot\delta\vec{E}_\text{com}$ 
(see the first equality in Eq.\ \ref{eq:app:dtI=div}).
With Eqs.~(\ref{eq:app:Bpt}), (\ref{eq:app:jPsi}), and (\ref{eq:app:fA-Psi}) 
we have, after some algebra,
\begin{equation}
\label{eq:app:EpxDphi}
c \delta\vec{E}_\text{com}^\pind\times\vnabla\varphi = -\frac{c^2 J_{ep}}{4\pi e^2 n_e^2} I'_\Psi (\vnabla\varphi)^2\vnabla\Psi + \frac{n_n^2}{n_b J_{np}} I (\vnabla\varphi)^2\vnabla\delta\mu_n - \frac{c(\vnabla\varphi)^2}{4\pi e n_e}\left( \grsh\Psi + I I'_\Psi \right)\vec{B}^\pind.
\end{equation}
Using again the equality $\diver\vec{B}^\pind = 0$,
we obtain for the first term in the right-hand side of Eq.~(\ref{eq:app:dtI=div})
\begin{multline}
\label{eq:app:div1}
-\diver\left( c\delta\vec{E}_\text{com}^\pind\times\vnabla\varphi \right) = \frac{c^2 J_{ep}}{4\pi e^2 n_e^2} (\vnabla\varphi)^2 I'_\Psi \grsh\Psi + (\vnabla\varphi)^2\vnabla \left( \frac{c^2 J_{ep}}{4\pi e^2 n_e^2} I'_\Psi \right) \cdot \vnabla\Psi - \frac{n_n^2}{n_b J_{np}}(\vnabla\varphi)^2 I'_\Psi \vnabla\delta\mu_n\cdot\vnabla\Psi -\\
- I \diver\left[ \frac{n_n^2}{n_b J_{np}}(\vnabla\varphi)^2\vnabla\delta\mu_n \right] + \vec{B}^\pind \cdot \vnabla\left[ \frac{c(\vnabla\varphi)^2}{4\pi e n_e} \left( \grsh\Psi + I I'_\Psi \right) \right].
\end{multline}
Plugging Eqs.~(\ref{eq:app:div3}), (\ref{eq:app:div2}), 
and (\ref{eq:app:div1}) into Eq.~(\ref{eq:app:dtI=div}) 
one can see that the second term in the right-hand side of Eq.~(\ref{eq:app:div2}) 
and the first and third terms in the right-hand side of Eq.~(\ref{eq:app:div1}) 
are combined to give 
$(\vnabla\varphi)^2 I'_\Psi \partial\Psi/\partial t$ 
[see Eq.~(\ref{eq:app:dtPsi-fin})]; 
this term then cancels a similar term in the left-hand side of Eq.~(\ref{eq:app:dtI=div}). 
Finally, one gets 
\begin{multline}
\label{eq:app:It}
(\vnabla\varphi)^2 I'_t = (\vnabla\varphi)^2 \vnabla\left( \frac{c^2 J_{ep}}{4\pi e^2 n_e^2} I'_\Psi \right) \cdot \vnabla\Psi - I \diver\left[ (\vnabla\varphi)^2 \left( \frac{n_n^2}{n_b J_{np}}\vnabla\delta\mu_n + \vec{U}_b^\pind \right) \right] +\\
+ \vec{B}^\pind \cdot \vnabla\left[ \frac{c(\vnabla\varphi)^2}{4\pi e n_e} \left( \grsh\Psi + I I'_\Psi \right) + \vnabla\varphi\cdot\vec{U}_b^\tind \right],
\end{multline}
which is the required Eq.\ (\ref{eq:It}).

\section{Spatial derivatives of unperturbed number densities near the crust-core interface}
\label{app:dndr}

The force balance equation
for an NS in hydrostatic equilibrium takes the form \cite{ov39,t39}  
\begin{equation}
\label{eq:app:TOV}
\frac{\diff P}{\diff r} = - G\, \frac{m\rho}{r^2} \, \frac{\left( 1 + \frac{P}{\rho c^2} \right)\left( 1 + \frac{4\pi r^3 P}{m c^2} \right)}{1 - \frac{2Gm}{r c^2}},
\end{equation}
where $m(r)$ is the gravitational mass enclosed in the sphere of radius $r$.
Since the mass of the crust does not exceed a few per cent of the total stellar mass, 
we can safely set $m\approx M$ in the vicinity of the crust-core interface. 

In the cold matter one has $P + \rho c^2 = n_b c^2 \diff\rho / \diff n_b$. 
Using this formula, one can rewrite the left-hand side of Eq.~(\ref{eq:app:TOV}) as
\begin{equation}
\label{eq:app:dPdr}
\frac{\diff P}{\diff r} = \frac{\diff P}{\diff\rho} \frac{\diff\rho}{\diff n_b} \frac{\diff n_b}{\diff r} = \left( \frac{c_s}{c} \right)^2 \frac{P+\rho c^2}{n_b} \frac{\diff n_b}{\diff r},
\end{equation}
where $c_s = \sqrt{\diff P/\diff\rho}$ is the equilibrium speed of sound.
Inserting Eq.~(\ref{eq:app:dPdr}) into (\ref{eq:app:TOV}),
one obtains
\begin{equation}
\label{eq:app:dnbdr}
\frac{\diff n_b / \diff r}{n_b/r}\Bigr|_{r\approx R_\text{core}} \approx -\frac{1}{2}\left( \frac{c}{c_s} \right)^2 \left( \frac{R_\text{core}c^2}{2GM} - 1 \right)^{-1},
\end{equation}
where we have neglected $P$ in comparison to $\rho c^2$,
which is justifiable at $n_b \sim n_0$ \cite{hpy07}.
Near the crust-core interface, at $n_b \sim 0.5n_0$, 
the ratio $(c/c_s)^2$, entering the right-hand side of Eq.\ (\ref{eq:app:dnbdr}), 
is much greater than 1. 
Deeper in the core EOS stiffens 
and the ratio $(c/c_s)^2$ becomes smaller, but remains several times greater than 1.
Note that the same result could be derived for a number density of any particle species.
Moreover, similar estimates can also be made for higher-order derivatives 
[e.g., $n_b''(r)/(n_b'/r)$], 
but the expressions will be more complicated, involving derivatives of $c_s$. 
However, this will only increase the effect.

Equation (\ref{eq:app:dnbdr}) shows that the standard estimate for the spatial derivative in the core, 
$\diff/\diff r \sim 1/R$ or $1/R_\text{core}$, 
is invalid for unperturbed quantities near the crust-core interface. 
In other words, a typical lengthscale in the outer layers of the core 
is significantly smaller than the radius, 
being of the order of the crust thickness.


\begin{thebibliography}{48}
	\expandafter\ifx\csname natexlab\endcsname\relax\def\natexlab#1{#1}\fi
	\expandafter\ifx\csname bibnamefont\endcsname\relax
	\def\bibnamefont#1{#1}\fi
	\expandafter\ifx\csname bibfnamefont\endcsname\relax
	\def\bibfnamefont#1{#1}\fi
	\expandafter\ifx\csname citenamefont\endcsname\relax
	\def\citenamefont#1{#1}\fi
	\expandafter\ifx\csname url\endcsname\relax
	\def\url#1{\texttt{#1}}\fi
	\expandafter\ifx\csname urlprefix\endcsname\relax\def\urlprefix{URL }\fi
	\providecommand{\bibinfo}[2]{#2}
	\providecommand{\eprint}[2][]{\url{#2}}
	
	\bibitem[{\citenamefont{{Jones}}(1988)}]{jones88}
	\bibinfo{author}{\bibfnamefont{P.~B.} \bibnamefont{{Jones}}},
	\bibinfo{journal}{\mnras} \textbf{\bibinfo{volume}{233}},
	\bibinfo{pages}{875} (\bibinfo{year}{1988}).
	
	\bibitem[{\citenamefont{{Shalybkov} and {Urpin}}(1997)}]{su97}
	\bibinfo{author}{\bibfnamefont{D.~A.} \bibnamefont{{Shalybkov}}}
	\bibnamefont{and} \bibinfo{author}{\bibfnamefont{V.~A.}
		\bibnamefont{{Urpin}}}, \bibinfo{journal}{\aap}
	\textbf{\bibinfo{volume}{321}}, \bibinfo{pages}{685} (\bibinfo{year}{1997}).
	
	\bibitem[{\citenamefont{{Rheinhardt} and {Geppert}}(2002)}]{rg02}
	\bibinfo{author}{\bibfnamefont{M.}~\bibnamefont{{Rheinhardt}}}
	\bibnamefont{and}
	\bibinfo{author}{\bibfnamefont{U.}~\bibnamefont{{Geppert}}},
	\bibinfo{journal}{Physical Review Letters} \textbf{\bibinfo{volume}{88}},
	\bibinfo{eid}{101103} (\bibinfo{year}{2002}).
	
	\bibitem[{\citenamefont{{Hollerbach} and {R{\"u}diger}}(2004)}]{hr04}
	\bibinfo{author}{\bibfnamefont{R.}~\bibnamefont{{Hollerbach}}}
	\bibnamefont{and}
	\bibinfo{author}{\bibfnamefont{G.}~\bibnamefont{{R{\"u}diger}}},
	\bibinfo{journal}{\mnras} \textbf{\bibinfo{volume}{347}},
	\bibinfo{pages}{1273} (\bibinfo{year}{2004}).
	
	\bibitem[{\citenamefont{{Gourgouliatos}
			et~al.}(2013)\citenamefont{{Gourgouliatos}, {Cumming}, {Reisenegger},
			{Armaza}, {Lyutikov}, and {Valdivia}}}]{gcr_etal13}
	\bibinfo{author}{\bibfnamefont{K.~N.} \bibnamefont{{Gourgouliatos}}},
	\bibinfo{author}{\bibfnamefont{A.}~\bibnamefont{{Cumming}}},
	\bibinfo{author}{\bibfnamefont{A.}~\bibnamefont{{Reisenegger}}},
	\bibinfo{author}{\bibfnamefont{C.}~\bibnamefont{{Armaza}}},
	\bibinfo{author}{\bibfnamefont{M.}~\bibnamefont{{Lyutikov}}},
	\bibnamefont{and} \bibinfo{author}{\bibfnamefont{J.~A.}
		\bibnamefont{{Valdivia}}}, \bibinfo{journal}{\mnras}
	\textbf{\bibinfo{volume}{434}}, \bibinfo{pages}{2480} (\bibinfo{year}{2013}).
	
	\bibitem[{\citenamefont{{Vigan{\`o}} et~al.}(2013)\citenamefont{{Vigan{\`o}},
			{Rea}, {Pons}, {Perna}, {Aguilera}, and {Miralles}}}]{vigano_etal13}
	\bibinfo{author}{\bibfnamefont{D.}~\bibnamefont{{Vigan{\`o}}}},
	\bibinfo{author}{\bibfnamefont{N.}~\bibnamefont{{Rea}}},
	\bibinfo{author}{\bibfnamefont{J.~A.} \bibnamefont{{Pons}}},
	\bibinfo{author}{\bibfnamefont{R.}~\bibnamefont{{Perna}}},
	\bibinfo{author}{\bibfnamefont{D.~N.} \bibnamefont{{Aguilera}}},
	\bibnamefont{and} \bibinfo{author}{\bibfnamefont{J.~A.}
		\bibnamefont{{Miralles}}}, \bibinfo{journal}{\mnras}
	\textbf{\bibinfo{volume}{434}}, \bibinfo{pages}{123} (\bibinfo{year}{2013}).
	
	\bibitem[{\citenamefont{{Gourgouliatos} and
			{Cumming}}(2014{\natexlab{a}})}]{gc14}
	\bibinfo{author}{\bibfnamefont{K.~N.} \bibnamefont{{Gourgouliatos}}}
	\bibnamefont{and}
	\bibinfo{author}{\bibfnamefont{A.}~\bibnamefont{{Cumming}}},
	\bibinfo{journal}{Physical Review Letters} \textbf{\bibinfo{volume}{112}},
	\bibinfo{eid}{171101} (\bibinfo{year}{2014}{\natexlab{a}}).
	
	\bibitem[{\citenamefont{{Gourgouliatos}
			et~al.}(2016)\citenamefont{{Gourgouliatos}, {Wood}, and
			{Hollerbach}}}]{gwh16}
	\bibinfo{author}{\bibfnamefont{K.~N.} \bibnamefont{{Gourgouliatos}}},
	\bibinfo{author}{\bibfnamefont{T.~S.} \bibnamefont{{Wood}}},
	\bibnamefont{and}
	\bibinfo{author}{\bibfnamefont{R.}~\bibnamefont{{Hollerbach}}},
	\bibinfo{journal}{Proceedings of the National Academy of Science}
	\textbf{\bibinfo{volume}{113}}, \bibinfo{pages}{3944} (\bibinfo{year}{2016}).
	
	\bibitem[{\citenamefont{{Elfritz} et~al.}(2016)\citenamefont{{Elfritz}, {Pons},
			{Rea}, {Glampedakis}, and {Vigan{\`o}}}}]{eprgv16}
	\bibinfo{author}{\bibfnamefont{J.~G.} \bibnamefont{{Elfritz}}},
	\bibinfo{author}{\bibfnamefont{J.~A.} \bibnamefont{{Pons}}},
	\bibinfo{author}{\bibfnamefont{N.}~\bibnamefont{{Rea}}},
	\bibinfo{author}{\bibfnamefont{K.}~\bibnamefont{{Glampedakis}}},
	\bibnamefont{and}
	\bibinfo{author}{\bibfnamefont{D.}~\bibnamefont{{Vigan{\`o}}}},
	\bibinfo{journal}{\mnras} \textbf{\bibinfo{volume}{456}},
	\bibinfo{pages}{4461} (\bibinfo{year}{2016}).
	
	\bibitem[{\citenamefont{{Bransgrove} et~al.}(2018)\citenamefont{{Bransgrove},
			{Levin}, and {Beloborodov}}}]{blb2017}
	\bibinfo{author}{\bibfnamefont{A.}~\bibnamefont{{Bransgrove}}},
	\bibinfo{author}{\bibfnamefont{Y.}~\bibnamefont{{Levin}}}, \bibnamefont{and}
	\bibinfo{author}{\bibfnamefont{A.}~\bibnamefont{{Beloborodov}}},
	\bibinfo{journal}{\mnras} \textbf{\bibinfo{volume}{473}},
	\bibinfo{pages}{2771} (\bibinfo{year}{2018}).
	
	\bibitem[{\citenamefont{{Passamonti} et~al.}(2017)\citenamefont{{Passamonti},
			{Akg{\"u}n}, {Pons}, and {Miralles}}}]{papm17}
	\bibinfo{author}{\bibfnamefont{A.}~\bibnamefont{{Passamonti}}},
	\bibinfo{author}{\bibfnamefont{T.}~\bibnamefont{{Akg{\"u}n}}},
	\bibinfo{author}{\bibfnamefont{J.~A.} \bibnamefont{{Pons}}},
	\bibnamefont{and} \bibinfo{author}{\bibfnamefont{J.~A.}
		\bibnamefont{{Miralles}}}, \bibinfo{journal}{\mnras}
	\textbf{\bibinfo{volume}{465}}, \bibinfo{pages}{3416} (\bibinfo{year}{2017}).
	
	\bibitem[{\citenamefont{{Castillo} et~al.}(2017)\citenamefont{{Castillo},
			{Reisenegger}, and {Valdivia}}}]{crv2017}
	\bibinfo{author}{\bibfnamefont{F.}~\bibnamefont{{Castillo}}},
	\bibinfo{author}{\bibfnamefont{A.}~\bibnamefont{{Reisenegger}}},
	\bibnamefont{and} \bibinfo{author}{\bibfnamefont{J.~A.}
		\bibnamefont{{Valdivia}}}, \bibinfo{journal}{\mnras}
	\textbf{\bibinfo{volume}{471}}, \bibinfo{pages}{507} (\bibinfo{year}{2017}).
	
	\bibitem[{\citenamefont{{Jones}}(2006)}]{jones06}
	\bibinfo{author}{\bibfnamefont{P.~B.} \bibnamefont{{Jones}}},
	\bibinfo{journal}{\mnras} \textbf{\bibinfo{volume}{365}},
	\bibinfo{pages}{339} (\bibinfo{year}{2006}), \eprint{astro-ph/0510396}.
	
	\bibitem[{\citenamefont{Gusakov et~al.}(2017)\citenamefont{Gusakov, Kantor, and
			Ofengeim}}]{GKO17}
	\bibinfo{author}{\bibfnamefont{M.~E.} \bibnamefont{Gusakov}},
	\bibinfo{author}{\bibfnamefont{E.~M.} \bibnamefont{Kantor}},
	\bibnamefont{and} \bibinfo{author}{\bibfnamefont{D.~D.}
		\bibnamefont{Ofengeim}}, \bibinfo{journal}{Phys. Rev. D}
	\textbf{\bibinfo{volume}{96}}, \bibinfo{pages}{103012}
	(\bibinfo{year}{2017}).
	
	\bibitem[{\citenamefont{{Goldreich} and {Reisenegger}}(1992)}]{gr92}
	\bibinfo{author}{\bibfnamefont{P.}~\bibnamefont{{Goldreich}}} \bibnamefont{and}
	\bibinfo{author}{\bibfnamefont{A.}~\bibnamefont{{Reisenegger}}},
	\bibinfo{journal}{\apj} \textbf{\bibinfo{volume}{395}}, \bibinfo{pages}{250}
	(\bibinfo{year}{1992}).
	
	\bibitem[{\citenamefont{{Thompson} and {Duncan}}(1996)}]{td96}
	\bibinfo{author}{\bibfnamefont{C.}~\bibnamefont{{Thompson}}} \bibnamefont{and}
	\bibinfo{author}{\bibfnamefont{R.~C.} \bibnamefont{{Duncan}}},
	\bibinfo{journal}{\apj} \textbf{\bibinfo{volume}{473}}, \bibinfo{pages}{322}
	(\bibinfo{year}{1996}).
	
	\bibitem[{\citenamefont{{Hoyos} et~al.}(2008)\citenamefont{{Hoyos},
			{Reisenegger}, and {Valdivia}}}]{hrv08}
	\bibinfo{author}{\bibfnamefont{J.}~\bibnamefont{{Hoyos}}},
	\bibinfo{author}{\bibfnamefont{A.}~\bibnamefont{{Reisenegger}}},
	\bibnamefont{and} \bibinfo{author}{\bibfnamefont{J.~A.}
		\bibnamefont{{Valdivia}}}, \bibinfo{journal}{\aap}
	\textbf{\bibinfo{volume}{487}}, \bibinfo{pages}{789} (\bibinfo{year}{2008}).
	
	\bibitem[{\citenamefont{{Reisenegger}}(2009)}]{reisenegger09}
	\bibinfo{author}{\bibfnamefont{A.}~\bibnamefont{{Reisenegger}}},
	\bibinfo{journal}{\aap} \textbf{\bibinfo{volume}{499}}, \bibinfo{pages}{557}
	(\bibinfo{year}{2009}).
	
	\bibitem[{\citenamefont{{Hoyos} et~al.}(2010)\citenamefont{{Hoyos},
			{Reisenegger}, and {Valdivia}}}]{hrv10}
	\bibinfo{author}{\bibfnamefont{J.~H.} \bibnamefont{{Hoyos}}},
	\bibinfo{author}{\bibfnamefont{A.}~\bibnamefont{{Reisenegger}}},
	\bibnamefont{and} \bibinfo{author}{\bibfnamefont{J.~A.}
		\bibnamefont{{Valdivia}}}, \bibinfo{journal}{\mnras}
	\textbf{\bibinfo{volume}{408}}, \bibinfo{pages}{1730} (\bibinfo{year}{2010}).
	
	\bibitem[{\citenamefont{{Glampedakis}
			et~al.}(2011{\natexlab{a}})\citenamefont{{Glampedakis}, {Jones}, and
			{Samuelsson}}}]{gjs11}
	\bibinfo{author}{\bibfnamefont{K.}~\bibnamefont{{Glampedakis}}},
	\bibinfo{author}{\bibfnamefont{D.~I.} \bibnamefont{{Jones}}},
	\bibnamefont{and}
	\bibinfo{author}{\bibfnamefont{L.}~\bibnamefont{{Samuelsson}}},
	\bibinfo{journal}{\mnras} \textbf{\bibinfo{volume}{413}},
	\bibinfo{pages}{2021} (\bibinfo{year}{2011}{\natexlab{a}}).
	
	\bibitem[{\citenamefont{{Beloborodov} and {Li}}(2016)}]{bl16}
	\bibinfo{author}{\bibfnamefont{A.~M.} \bibnamefont{{Beloborodov}}}
	\bibnamefont{and} \bibinfo{author}{\bibfnamefont{X.}~\bibnamefont{{Li}}},
	\bibinfo{journal}{\apj} \textbf{\bibinfo{volume}{833}}, \bibinfo{eid}{261}
	(\bibinfo{year}{2016}).
	
	\bibitem[{\citenamefont{{Sinha} and {Sedrakian}}(2015)}]{ss15}
	\bibinfo{author}{\bibfnamefont{M.}~\bibnamefont{{Sinha}}} \bibnamefont{and}
	\bibinfo{author}{\bibfnamefont{A.}~\bibnamefont{{Sedrakian}}},
	\bibinfo{journal}{\prc} \textbf{\bibinfo{volume}{91}}, \bibinfo{eid}{035805}
	(\bibinfo{year}{2015}), \eprint{1502.02979}.
	
	\bibitem[{\citenamefont{{Sedrakian} et~al.}(2017)\citenamefont{{Sedrakian},
			{Xu-Guang}, {Sinha}, and {Clark}}}]{sxsc17}
	\bibinfo{author}{\bibfnamefont{A.}~\bibnamefont{{Sedrakian}}},
	\bibinfo{author}{\bibfnamefont{H.}~\bibnamefont{{Xu-Guang}}},
	\bibinfo{author}{\bibfnamefont{M.}~\bibnamefont{{Sinha}}}, \bibnamefont{and}
	\bibinfo{author}{\bibfnamefont{J.~W.} \bibnamefont{{Clark}}}, in
	\emph{\bibinfo{booktitle}{Journal of Physics Conference Series}}
	(\bibinfo{year}{2017}), vol. \bibinfo{volume}{861} of
	\emph{\bibinfo{series}{Journal of Physics Conference Series}}, p.
	\bibinfo{pages}{012025}, \eprint{1701.00895}.
	
	\bibitem[{\citenamefont{{Iakovlev} and {Shalybkov}}(1991)}]{ys91a}
	\bibinfo{author}{\bibfnamefont{D.~G.} \bibnamefont{{Iakovlev}}}
	\bibnamefont{and} \bibinfo{author}{\bibfnamefont{D.~A.}
		\bibnamefont{{Shalybkov}}}, \bibinfo{journal}{Astrophys. Sp. Sci.}
	\textbf{\bibinfo{volume}{176}}, \bibinfo{pages}{171} (\bibinfo{year}{1991}).
	
	\bibitem[{\citenamefont{{Reisenegger}}(1995{\natexlab{a}})}]{Reis95}
	\bibinfo{author}{\bibfnamefont{A.}~\bibnamefont{{Reisenegger}}},
	\bibinfo{journal}{\apj} \textbf{\bibinfo{volume}{442}}, \bibinfo{pages}{749}
	(\bibinfo{year}{1995}{\natexlab{a}}).
	
	\bibitem[{\citenamefont{{Haensel} et~al.}(2007)\citenamefont{{Haensel},
			{Potekhin}, and {Yakovlev}}}]{hpy07}
	\bibinfo{editor}{\bibfnamefont{P.}~\bibnamefont{{Haensel}}},
	\bibinfo{editor}{\bibfnamefont{A.~Y.} \bibnamefont{{Potekhin}}},
	\bibnamefont{and} \bibinfo{editor}{\bibfnamefont{D.~G.}
		\bibnamefont{{Yakovlev}}}, eds., \emph{\bibinfo{title}{{Neutron Stars 1 :
				Equation of State and Structure}}}, vol. \bibinfo{volume}{326} of
	\emph{\bibinfo{series}{Astrophysics and Space Science Library}}
	(\bibinfo{year}{2007}).
	
	\bibitem[{\citenamefont{{Cowling}}(1941)}]{cow41}
	\bibinfo{author}{\bibfnamefont{T.~G.} \bibnamefont{{Cowling}}},
	\bibinfo{journal}{\mnras} \textbf{\bibinfo{volume}{101}},
	\bibinfo{pages}{367} (\bibinfo{year}{1941}).
	
	\bibitem[{\citenamefont{{Shalybkov} and {Urpin}}(1995)}]{su95}
	\bibinfo{author}{\bibfnamefont{D.~A.} \bibnamefont{{Shalybkov}}}
	\bibnamefont{and} \bibinfo{author}{\bibfnamefont{V.~A.}
		\bibnamefont{{Urpin}}}, \bibinfo{journal}{\mnras}
	\textbf{\bibinfo{volume}{273}}, \bibinfo{pages}{643} (\bibinfo{year}{1995}).
	
	\bibitem[{\citenamefont{{Goedbloed} et~al.}(2010)\citenamefont{{Goedbloed},
			{Keppens}, and {Poedts}}}]{Goed2010}
	\bibinfo{author}{\bibfnamefont{J.~P.} \bibnamefont{{Goedbloed}}},
	\bibinfo{author}{\bibfnamefont{R.}~\bibnamefont{{Keppens}}},
	\bibnamefont{and} \bibinfo{author}{\bibfnamefont{S.}~\bibnamefont{{Poedts}}},
	\emph{\bibinfo{title}{{Advanced Magnetohydrodynamics}}}
	(\bibinfo{publisher}{Cambridge University Press, Cambridge, UK},
	\bibinfo{year}{2010}).
	
	\bibitem[{\citenamefont{{Yakovlev} et~al.}(2001)\citenamefont{{Yakovlev},
			{Kaminker}, {Gnedin}, and {Haensel}}}]{ykgh01}
	\bibinfo{author}{\bibfnamefont{D.~G.} \bibnamefont{{Yakovlev}}},
	\bibinfo{author}{\bibfnamefont{A.~D.} \bibnamefont{{Kaminker}}},
	\bibinfo{author}{\bibfnamefont{O.~Y.} \bibnamefont{{Gnedin}}},
	\bibnamefont{and}
	\bibinfo{author}{\bibfnamefont{P.}~\bibnamefont{{Haensel}}},
	\bibinfo{journal}{\physrep} \textbf{\bibinfo{volume}{354}},
	\bibinfo{pages}{1} (\bibinfo{year}{2001}).
	
	\bibitem[{\citenamefont{{Yakovlev} and {Shalybkov}}(1990)}]{ys90}
	\bibinfo{author}{\bibfnamefont{D.~G.} \bibnamefont{{Yakovlev}}}
	\bibnamefont{and} \bibinfo{author}{\bibfnamefont{D.~A.}
		\bibnamefont{{Shalybkov}}}, \bibinfo{journal}{Soviet Astronomy Letters}
	\textbf{\bibinfo{volume}{16}}, \bibinfo{pages}{86} (\bibinfo{year}{1990}).
	
	\bibitem[{\citenamefont{{Gourgouliatos} and
			{Cumming}}(2014{\natexlab{b}})}]{GourCum14}
	\bibinfo{author}{\bibfnamefont{K.~N.} \bibnamefont{{Gourgouliatos}}}
	\bibnamefont{and}
	\bibinfo{author}{\bibfnamefont{A.}~\bibnamefont{{Cumming}}},
	\bibinfo{journal}{\mnras} \textbf{\bibinfo{volume}{438}},
	\bibinfo{pages}{1618} (\bibinfo{year}{2014}{\natexlab{b}}),
	\eprint{1311.7004}.
	
	\bibitem[{\citenamefont{{Beskin} et~al.}(2006)\citenamefont{{Beskin},
			{Gurevich}, and {Istomin}}}]{BGI2006}
	\bibinfo{author}{\bibfnamefont{V.~S.} \bibnamefont{{Beskin}}},
	\bibinfo{author}{\bibfnamefont{A.~V.} \bibnamefont{{Gurevich}}},
	\bibnamefont{and} \bibinfo{author}{\bibfnamefont{Y.~N.}
		\bibnamefont{{Istomin}}}, \emph{\bibinfo{title}{{Physics of the Pulsar
				Magnetosphere}}} (\bibinfo{publisher}{Cambridge University Press, Cambridge,
		UK}, \bibinfo{year}{2006}).
	
	\bibitem[{\citenamefont{{Glampedakis} and {Lasky}}(2015)}]{gl15}
	\bibinfo{author}{\bibfnamefont{K.}~\bibnamefont{{Glampedakis}}}
	\bibnamefont{and} \bibinfo{author}{\bibfnamefont{P.~D.}
		\bibnamefont{{Lasky}}}, \bibinfo{journal}{\mnras}
	\textbf{\bibinfo{volume}{450}}, \bibinfo{pages}{1638} (\bibinfo{year}{2015}),
	\eprint{1501.05473}.
	
	\bibitem[{\citenamefont{{Heiselberg} and {Hjorth-Jensen}}(1999)}]{hhj99}
	\bibinfo{author}{\bibfnamefont{H.}~\bibnamefont{{Heiselberg}}}
	\bibnamefont{and}
	\bibinfo{author}{\bibfnamefont{M.}~\bibnamefont{{Hjorth-Jensen}}},
	\bibinfo{journal}{\apjl} \textbf{\bibinfo{volume}{525}}, \bibinfo{pages}{L45}
	(\bibinfo{year}{1999}).
	
	\bibitem[{\citenamefont{{Oppenheimer} and {Volkoff}}(1939)}]{ov39}
	\bibinfo{author}{\bibfnamefont{J.~R.} \bibnamefont{{Oppenheimer}}}
	\bibnamefont{and} \bibinfo{author}{\bibfnamefont{G.~M.}
		\bibnamefont{{Volkoff}}}, \bibinfo{journal}{Physical Review}
	\textbf{\bibinfo{volume}{55}}, \bibinfo{pages}{374} (\bibinfo{year}{1939}).
	
	\bibitem[{\citenamefont{{Tolman}}(1939)}]{t39}
	\bibinfo{author}{\bibfnamefont{R.~C.} \bibnamefont{{Tolman}}},
	\bibinfo{journal}{Physical Review} \textbf{\bibinfo{volume}{55}},
	\bibinfo{pages}{364} (\bibinfo{year}{1939}).
	
	\bibitem[{\citenamefont{{Kaminker} et~al.}(2016)\citenamefont{{Kaminker},
			{Yakovlev}, and {Haensel}}}]{kyh16}
	\bibinfo{author}{\bibfnamefont{A.~D.} \bibnamefont{{Kaminker}}},
	\bibinfo{author}{\bibfnamefont{D.~G.} \bibnamefont{{Yakovlev}}},
	\bibnamefont{and}
	\bibinfo{author}{\bibfnamefont{P.}~\bibnamefont{{Haensel}}},
	\bibinfo{journal}{\apss} \textbf{\bibinfo{volume}{361}}, \bibinfo{eid}{267}
	(\bibinfo{year}{2016}), \eprint{1607.05265}.
	
	\bibitem[{\citenamefont{{Potekhin} et~al.}(2015)\citenamefont{{Potekhin},
			{Pons}, and {Page}}}]{ppp15}
	\bibinfo{author}{\bibfnamefont{A.~Y.} \bibnamefont{{Potekhin}}},
	\bibinfo{author}{\bibfnamefont{J.~A.} \bibnamefont{{Pons}}},
	\bibnamefont{and} \bibinfo{author}{\bibfnamefont{D.}~\bibnamefont{{Page}}},
	\bibinfo{journal}{\ssr} \textbf{\bibinfo{volume}{191}}, \bibinfo{pages}{239}
	(\bibinfo{year}{2015}), \eprint{1507.06186}.
	
	\bibitem[{\citenamefont{{Akg{\"u}n} et~al.}(2013)\citenamefont{{Akg{\"u}n},
			{Reisenegger}, {Mastrano}, and {Marchant}}}]{armm13}
	\bibinfo{author}{\bibfnamefont{T.}~\bibnamefont{{Akg{\"u}n}}},
	\bibinfo{author}{\bibfnamefont{A.}~\bibnamefont{{Reisenegger}}},
	\bibinfo{author}{\bibfnamefont{A.}~\bibnamefont{{Mastrano}}},
	\bibnamefont{and}
	\bibinfo{author}{\bibfnamefont{P.}~\bibnamefont{{Marchant}}},
	\bibinfo{journal}{\mnras} \textbf{\bibinfo{volume}{433}},
	\bibinfo{pages}{2445} (\bibinfo{year}{2013}), \eprint{1302.0273}.
	
	\bibitem[{\citenamefont{{Reisenegger}}(1995{\natexlab{b}})}]{reisenegger95}
	\bibinfo{author}{\bibfnamefont{A.}~\bibnamefont{{Reisenegger}}},
	\bibinfo{journal}{\apj} \textbf{\bibinfo{volume}{442}}, \bibinfo{pages}{749}
	(\bibinfo{year}{1995}{\natexlab{b}}), \eprint{astro-ph/9410035}.
	
	\bibitem[{\citenamefont{{Glampedakis}
			et~al.}(2011{\natexlab{b}})\citenamefont{{Glampedakis}, {Andersson}, and
			{Samuelsson}}}]{gas11}
	\bibinfo{author}{\bibfnamefont{K.}~\bibnamefont{{Glampedakis}}},
	\bibinfo{author}{\bibfnamefont{N.}~\bibnamefont{{Andersson}}},
	\bibnamefont{and}
	\bibinfo{author}{\bibfnamefont{L.}~\bibnamefont{{Samuelsson}}},
	\bibinfo{journal}{\mnras} \textbf{\bibinfo{volume}{410}},
	\bibinfo{pages}{805} (\bibinfo{year}{2011}{\natexlab{b}}),
	\eprint{1001.4046}.
	
	\bibitem[{\citenamefont{{Gusakov} and {Dommes}}(2016)}]{gd16}
	\bibinfo{author}{\bibfnamefont{M.~E.} \bibnamefont{{Gusakov}}}
	\bibnamefont{and} \bibinfo{author}{\bibfnamefont{V.~A.}
		\bibnamefont{{Dommes}}}, \bibinfo{journal}{\prd}
	\textbf{\bibinfo{volume}{94}}, \bibinfo{eid}{083006} (\bibinfo{year}{2016}).
	
	\bibitem[{\citenamefont{{Graber} et~al.}(2015)\citenamefont{{Graber},
			{Andersson}, {Glampedakis}, and {Lander}}}]{gagl15}
	\bibinfo{author}{\bibfnamefont{V.}~\bibnamefont{{Graber}}},
	\bibinfo{author}{\bibfnamefont{N.}~\bibnamefont{{Andersson}}},
	\bibinfo{author}{\bibfnamefont{K.}~\bibnamefont{{Glampedakis}}},
	\bibnamefont{and} \bibinfo{author}{\bibfnamefont{S.~K.}
		\bibnamefont{{Lander}}}, \bibinfo{journal}{\mnras}
	\textbf{\bibinfo{volume}{453}}, \bibinfo{pages}{671} (\bibinfo{year}{2015}).
	
	\bibitem[{\citenamefont{{Olausen} and {Kaspi}}(2014)}]{ok14}
	\bibinfo{author}{\bibfnamefont{S.~A.} \bibnamefont{{Olausen}}}
	\bibnamefont{and} \bibinfo{author}{\bibfnamefont{V.~M.}
		\bibnamefont{{Kaspi}}}, \bibinfo{journal}{\apjs}
	\textbf{\bibinfo{volume}{212}}, \bibinfo{eid}{6} (\bibinfo{year}{2014}),
	\eprint{1309.4167}.
	
	\bibitem[{\citenamefont{{Popov} and {Postnov}}(2013)}]{pp13}
	\bibinfo{author}{\bibfnamefont{S.~B.} \bibnamefont{{Popov}}} \bibnamefont{and}
	\bibinfo{author}{\bibfnamefont{K.~A.} \bibnamefont{{Postnov}}},
	\bibinfo{journal}{ArXiv e-prints}  (\bibinfo{year}{2013}),
	\eprint{1307.4924}.
	
	\bibitem[{\citenamefont{{Beloborodov}}(2017)}]{beloborodov17}
	\bibinfo{author}{\bibfnamefont{A.~M.} \bibnamefont{{Beloborodov}}},
	\bibinfo{journal}{\apjl} \textbf{\bibinfo{volume}{843}}, \bibinfo{eid}{L26}
	(\bibinfo{year}{2017}), \eprint{1702.08644}.
	
	\bibitem[{\citenamefont{{Popov} et~al.}(2018)\citenamefont{{Popov}, {Postnov},
			and {Pshirkov}}}]{ppp18}
	\bibinfo{author}{\bibfnamefont{S.~B.} \bibnamefont{{Popov}}},
	\bibinfo{author}{\bibfnamefont{K.~A.} \bibnamefont{{Postnov}}},
	\bibnamefont{and} \bibinfo{author}{\bibfnamefont{M.~S.}
		\bibnamefont{{Pshirkov}}}, \bibinfo{journal}{ArXiv e-prints}
	(\bibinfo{year}{2018}), \eprint{1801.00640}.
	
\end{thebibliography}

\end{document}